\documentclass[a4paper,12pt]{article}
\usepackage[margin=1in]{geometry}
\usepackage{url}

\usepackage{tabularx}
\usepackage{booktabs}
\usepackage{graphicx}
\usepackage{soul}

\usepackage[utf8]{inputenc}
\usepackage{enumitem}
\usepackage{hyperref}

\usepackage[most]{tcolorbox}
\tcbuselibrary{breakable}
\usepackage{enumitem}

\usepackage{algorithm}
\usepackage{algpseudocode}
\usepackage{xcolor}
\usepackage{listings}

\definecolor{codegreen}{rgb}{0,0.6,0}
\definecolor{codegray}{rgb}{0.5,0.5,0.5}
\definecolor{codepurple}{rgb}{0.58,0,0.82}
\definecolor{backcolour}{rgb}{0.95,0.95,0.92}

\lstdefinestyle{mystyle}{
    backgroundcolor=\color{backcolour},
    commentstyle=\color{codegreen},
    keywordstyle=\color{magenta},
    numberstyle=\tiny\color{codegray},
    stringstyle=\color{codepurple},
    basicstyle=\ttfamily\footnotesize,
    breaklines=true,
    showstringspaces=false
}

\usepackage{apacite}
\usepackage{natbib}

\usepackage{xcolor}

\usepackage{makecell}

\usepackage{float}

\usepackage[normalem]{ulem}

\usepackage{threeparttable} 

\usepackage{tikz}
\usetikzlibrary{arrows.meta, positioning}
\tikzset{
  block/.style = {
    draw,
    rectangle,
    minimum width=3cm,
    minimum height=1cm,
    align=center
  },
  arrow/.style = {
    -{Latex[length=3mm]},
    thick
  }
}

\makeatletter
\renewcommand{\footnotesize}{\scriptsize}
\renewcommand{\@makefntext}[1]{%
  \noindent\makebox[1.8em][r]{\@thefnmark.}\ #1}
\makeatother

\title{Can large language models assist choice modelling? Insights into prompting strategies and current models' capabilities}

\author{%
  Georges Sfeir\thanks{Choice Modelling Centre \& Institute for Transport Studies, University of Leeds, Leeds, United Kingdom, G.Sfeir@leeds.ac.uk. Corresponding Author} \and
  Gabriel Nova\thanks{CityAI Lab, Transport and Logistics group, Engineering Systems and Services Department, TU Delft, Delft, Netherlands, G.Nova@tudelft.nl} \and
  Stephane Hess\thanks{Choice Modelling Centre \& Institute for Transport Studies, University of Leeds, Leeds, United Kingdom, S.Hess@leeds.ac.uk} \and 
  Sander van Cranenburgh\thanks{CityAI Lab, Transport and Logistics group, Engineering Systems and Services Department, TU Delft, Delft, Netherlands, S.vanCranenburgh@tudelft.nl}}

\date{}

\begin{document}

\maketitle
\vspace{-1cm}
\section*{Abstract}
Large Language Models (LLMs) are becoming widely used to support various workflows across different disciplines, yet their potential in discrete choice modelling remains relatively unexplored. This work examines the potential of LLMs as assistive agents in the specification and, where technically feasible, estimation of Multinomial Logit models. We implement a systematic experimental framework involving twelve versions of seven leading LLMs (ChatGPT, Claude, DeepSeek, Gemini, Gemma, Llama, and Mistral) evaluated under five experimental configurations. These configurations vary along three dimensions: (i) modelling goal (suggesting vs. suggesting and estimating MNL models); (ii) prompting strategy (Zero-Shot vs. Chain-of-Thoughts (CoT)); and (iii) information availability (full dataset vs. data dictionary summarising variable names and types). Each specification suggested by the LLMs is implemented, estimated, and evaluated based on goodness-of-fit metrics, behavioural plausibility, and model complexity. Our findings reveal that proprietary LLMs can generate valid and behaviourally sound utility specifications, particularly when guided by structured prompts (CoT). Open-weight models such as Llama and Gemma struggled to produce meaningful specifications. Notably, some LLMs performed better when provided with just data dictionary, suggesting that limiting raw data access may enhance internal reasoning capabilities. Among all LLMs, GPT o3, operating in an agentic setting, was uniquely capable of correctly estimating its own specifications by executing self-generated code. Overall, the results demonstrate both the promise and current limitations of LLMs as assistive agents in discrete choice modelling, not only for model specification but also for supporting modelling decision and estimation, and provide practical guidance for integrating these tools into choice modellers' workflows.

\vspace{0.3cm}
\enlargethispage{1.5\baselineskip}  
\textbf{Keywords}: Large Language Models, Discrete Choice Models,  Utility Specification
\clearpage
\section{Introduction}
Discrete Choice Models (DCMs), are widely used across numerous fields for understanding and analysing individual choice behaviour \citep{hess2024handbook, mariel2021environmental,liebe2023maximizing,buckell2022utility}. Modellers use these structures to mathematically represent decision-makers’ observed choices, estimate parameters that capture their current preferences, and apply the models to forecast demand in future scenarios.\\ 

\noindent The process of conducting a choice modelling study involves numerous modeller decisions, drawing on behavioural theories, data constraints, and researcher assumptions \citep[cf.][]{nova2024understanding, van2022choice}. Modellers need to make a decision on model family (e.g. random utility \emph{vs} preference accumulation) and then choose a specific structure within that model family (e.g. multinomial \emph{vs} nested logit) \citep{hessdalybatley2018}. They have to construct functional forms that map observed data, typically including alternative‐specific attributes (e.g. travel time, travel cost), individual‐specific socioeconomic and contextual variables (e.g. gender, age, income, region, weather), and alternative availabilities, into a formal definition of a value function such as utility, which includes decisions on linearity \emph{vs} non-linearity, as well as interactions. Modellers often propose a candidate specification, estimate its parameters, examine both goodness‐of‐fit statistics and theoretical plausibility, and then refine the specification, balancing parsimony, model fit, and behavioural assumptions. Even with a small set of variables, the space of possible specifications grows combinatorially, and despite decades of modelling advancements, specifying discrete choice models remains a time-intensive and expertise-driven task, requiring careful consideration of behavioural theory, statistical diagnostics, and domain knowledge.\\

\noindent Against this backdrop, the present paper explores whether the exponential growth in the development and application of Large Language Models (LLMs) across a wide range of tasks provides a novel opportunity in choice modelling. 
LLMs have proven themselves as powerful tools for queries that require understanding, generating, and reasoning technical text \citep{qu2025tool}. Moreover, recent work has demonstrated that LLMs achieve high performance on a range of tasks. For instance, GPT‐4 reaches near‐human accuracy on complex reasoning benchmarks and solves most grade‐school math problems \citep{wei2022chain}, while Claude 3 Opus and DeepSeek‐V3 similarly excel on mathematical and coding problems \citep{claude37sonnet,liu2024deepseek}. Although LLMs can parse academic literature, generate code, and even run them based on user prompts across diverse fields \citep{brown2020language, mialon2023augmented, wang2024survey}, their actual potential to provide assistance in the choice model specification process has not been explored.\\ 

\noindent This paper evaluates how current LLMs can assist in the specification of and, where technically feasible, the estimation of discrete choice models. Our aim is twofold. First, we seek to understand how modellers should interact with LLMs in an out-of-the-box manner (i.e. without additional LLM training or fine-tuning) to obtain useful and behaviourally plausible specifications. Second, we assess how different LLMs perform under these conditions, without claiming definitive rankings.
To this end, we design a set of experiments that vary by prompt type (i.e. the instruction given to the LLM), information provided to the LLM (full data vs. data dictionary only), and modelling task (specification vs. estimation). These variations allow us to identify interaction patterns that influence the quality and plausibility of LLM-generated specifications. Moreover, by varying the modelling task, we not only evaluate LLMs as assistive agents for specification, but also provide insights into their ability to conduct end-to-end econometric modelling, including the execution of self-generated code for estimation. We then apply this framework to twelve versions drawn from six prominent LLM families, including OpenAI’s GPT variants, Anthropic’s Claude series, DeepSeek Chat, Google’s Gemini and Gemma, Meta's Llama series, and Mistral le Chat. These models constitute a representative sample of widely used LLMs at the time of the study, rather than an exhaustive benchmark. All LLM-generated specifications are then evaluated based on behavioural plausibility (e.g. expected parameter signs), goodness-of-fit measures, and overall specification quality, as well as their robustness across experiment configurations (i.e. whether the same LLM consistently generate good specifications) and convergence reliability (i.e. whether generated specifications converge when implemented and reported estimates can be reproduced). By bridging the fields of DCM and AI, this study provides, to our knowledge, a first proof-of-concept assessment of the potential role of LLMs in behavioural econometric model development. The results not only offer practical insights for researchers seeking to augment their workflows with LLMs but also raise broader questions about the evolving boundary between human expertise and machine intelligence in scientific modelling. Of course, the development of LLMs continues apace, and the models applied in this paper are only current at the time of writing this paper--the broader points remain valid going forward.\\

\noindent The remainder of this paper is organized as follows. Section \ref{section:lit} reviews relevant literature at the intersection of DCMs and LLMs. Section \ref{section:met} describes the research methodology and presents the different LLMs, prompts, and dataset used in this study. Section \ref{section:res} reports and analyses the results across all experimental configurations. Finally, Section \ref{section:con} concludes with a discussion of key findings, limitations, and directions for future research.

\section{Literature}
\label{section:lit}
\noindent Model specification is the main phase of the choice modelling process, characterised by a semi-structured, iterative, demanding workflow that is often time consuming. Rooted in behavioural theories and guided by the modeller’s knowledge and prior assumptions, this task involves making decisions on model structure and specification of the value functions (e.g. utility functions) \citep{van2022choice,nova2025improving}. Modelling decisions include choices on the model family and error structure, which attributes to include, how to specify their functional form (e.g., linear, transformed, or interacted either among themselves or with covariates), and how to account for heterogeneity. Even with a modest number of variables, the space of possible model specification expands combinatorially \citep{rodrigues2024model}. In this section, we look at existing work on assisted model specification before turning our attention to LLMs. We then focus on LLM capabilities and their performance, which are shaped by prompting strategies and external tool accessibility.

\subsection{Existing work on assisted specification process}

\noindent To support or automate the model-building phase, recent research has explored a range of approaches, from combinatorial optimisation techniques to machine learning. Initially, researchers used static metaheuristics, such as Simulated Annealing, Bayesian variable selection, neighbourhood search, and bi-level optimisation, to explore the model space iteratively, often incorporating constraints to enhance behavioural plausibility and model fit \citep{paez2022discrete, rodrigues2020bayesian, ortelli2021assisted, beeramoole2023extensive}. More recently, grammar-based approaches have emerged, with Grammatical Evolution used either independently \citep{haj5195530grammar} or in bi-level frameworks to improve predictive consistency \citep{ghorbani2025enhanced}. Lastly, machine learning methods have also been used towards automating this process. For instance, \cite{mesbah5226710shap} used SHapley Additive exPlanations (SHAP) to inform the construction of behaviourally sound utility functions.  Building on these trends, \cite{nova2025improving} proposed a reinforcement learning framework that formulates model specification as an adaptative learning process rather than a static one. In their approach, a Deep Q-Network agent learns to propose model candidates through sequential interaction, guided by a reward function that encodes behavioural expectations and modelling outcomes.\\

\noindent However, existing assisted specification approaches typically represent domain knowledge through problem specific constraints, grammars, or reward signals. More generally, these approaches offer limited support for leveraging modelling knowledge, such as translating hypotheses or complex behavioural expectations into utility terms to ensure plausible specifications. As a result, the search process is tied to a particular formulation and often requires redesign when the dataset or modelling objective changes.

\subsection{Large Language Models }

Large language models are generative artificial intelligence models designed to process and generate human‐like text by learning lexical representation from large corpora \citep{naveed2023comprehensive}. They are typically trained with self-supervised objectives (e.g. next token prediction), which enables them to encode lexical elements and world knowledge, and later adapted to follow user instructions via fine-tuning \cite{chang2024survey}. This was achieved by leveraging the Transformer architecture, which uses multi-head self-attention to model distant relationships between tokens while enabling efficient parallel training. This design made it practical to scale both model size and training data, supporting transferability to new tasks\citep{vaswani2017attention, radford2019language, brown2020language}. Since then, LLMs have demonstrated increasing capabilities from text generation to mathematical problem-solving, coding, and reasoning.\\

\noindent As LLM capabilities have grown, they have been increasingly used not only as text generators but also embedded within pipeline-based systems \citep{yao2022react, openai2023functioncalling}. This shift means that the same underlying model can support different system behaviours. In a non-agentic setup, the model maps a prompt to an output with no access to external environments, which is effective for drafting, explaining concepts, and generating code when the necessary inputs are already provided \citep{chen2021evaluating,zhang2023instruction}. By contrast, agentic systems extend the model by enabling it to handle external tools to retrieve information, evaluate intermediate steps, and perform actions (e.g. search, code execution, APIs). While these systems can improve performance on data-intensive or multi-step tasks, their overall effectiveness depends not only on the model itself but also on tool availability, interfaces, permissions, and coordination overhead \citep{wang2024survey, schick2023toolformer}.\\

\noindent Finally, access to LLMs differs across the system. Closed-weight models are typically provided via proprietary APIs or interfaces, which simplifies deployment and allows non-technical users to interact with them directly. However, they limit access to internal parameters and training data, preventing users from fine-tuning or deploying the model on their own hardware. By contrast, open-weight models provide full access to trained weights, allow architectural replication, and support fine-tuning on local infrastructure, which enables integrate them into user’s systems. Although, the original training data and/or training code are not always disclosed. 
Recent examples include closed-weight families such as OpenAI and Anthropic, and open-weight families such as Meta’s Llama, Google’s Gemma, and Le Chat’s Mistral, and DeepSeek’s MoE models.

\subsection{Prompting strategies}\label{prompting_strategies}
A prompt is a structured instruction or query provided to a LLM to guide the form, content, or behaviour of its response. In general, a prompt consists of textual input, contextual information, system role (e.g. defining the model as a helpful assistant, a tutor, etc.), or potentially example-based demonstrations, which define the task the model is expected to perform and/or specify the format the response should follow. Some LLMs are capable of handling multimodal inputs, such as images, audio, files, or even code. \\

\noindent To be able to interpret prompts, LLMs decompose them into discrete units known as tokens, which are the basic units for computation. These tokens are mapped into high-dimensional latent spaces and passed through multiple layers of transformer-based neural architectures that predict the next most likely tokens in the output sequence. Consequently, the quality and structure of the input prompt highly influence the relevance, coherence, and robustness of the model’s output \citep{jin2020bert, minaee2021deep}. Moreover, without adequate context, these LLMs can also manifest undesirable behaviours \citep{bommasani2021opportunities}, biased responses \citep{gehman2020realtoxicityprompts}, provide false or incorrect information, or even generate hallucinatory content \citep{welleck2019neural}.\\

\noindent The quality, relevance, and factual consistency of model responses can be significantly impacted by structure, content, and context of prompts. For example, \cite{kojima2022large} show that models benefit from prompts that contains chain-of-thought instructions, enabling them to outperform the same prompt-task. Similarly, \cite{min2022rethinking} provide evidence that prompt context—including examples, step-by-step reasoning patterns, or structural templates—plays a more critical role than the prompt length or wording alone. This process, referred to as prompt engineering nowadays, is the practice of designing input prompts that effectively obtain accurate and context-aware responses from generative AI systems \citep{reynolds2021prompt, brown2020language}. \\

\noindent Prompt engineering can therefore take many forms depending on the degree of structure, domain knowledge, and contextual information integrated into the input. The two most extensively used prompting paradigms are Zero-Shot Prompting (ZSP) and Chain-of-Thought (CoT) prompting. ZSP relies solely on the model’s pre-trained knowledge and requires no additional context, examples, or step-by-step instructions \citep{brown2020language}. Although ZSP is versatile and straightforward to use, it often generates generic responses that lack detailed thoughts, especially in tasks that require specific domain knowledge. In contrast, CoT prompting guides the model through intermediate reasoning steps, encouraging more structured and coherent outputs \citep{wei2022chain}. It has been shown to significantly improve performance on tasks involving logical inference, arithmetic, and structured decision-making \citep{zhou2022least}. Beyond ZSP and CoT prompting, a commonly used paradigm is few-shot prompting, in which the model is provided with one or more examples that demonstrate how the task should be performed. Few-shot prompting, sometimes combined with chain-of-thought reasoning, has been shown to substantially improve performance on a wide range of tasks by enabling in-context learning and demonstration-based adaptation \citep{brown2020language, wei2022chain}. A key contribution of the present paper is to investigate the importance of the type of prompt used when deploying LLMs for choice modelling.

\section{Research methodology and empirical setup}
\label{section:met}
This section outlines the experimental framework used to evaluate the ability of LLMs to specify and, where technically feasible, estimate MNL models. In the present work, we limit our focus to MNL models only as the work serves as a proof of concept. Extensions to more complex specifications remain an avenue for future work. We start by looking at the overall methodological framework before looking at the specific context of our application.

\subsection{General framework}\label{exp_framework}

Our framework evaluates the performance of LLMs using two prompting strategies, zero-shot and chain-of-thought, while few-shot prompting is intentionally excluded from the present study. Our objective is not to maximise LLM performance through in-context or demonstration-based learning, but to evaluate how LLMs perform as assistive tools when used in an out-of-the-box manner, without curated examples or expert-provided templates. In the context of discrete choice modelling, few-shot prompting would require supplying example utility specifications and modelling decisions, thereby embedding domain expertise and behavioural assumptions directly into the prompt. This would confound the assessment of the LLMs’ intrinsic reasoning and specification capabilities. Moreover, most applied researchers and practitioners currently interact with LLMs using zero-shot or lightly structured prompts rather than creating few-shot prompts. We therefore focus on ZSP and CoT prompting to isolate the role of reasoning guidance while preserving external validity.\\

\noindent In our context, the ZSP template asked the LLM to suggest and estimate (where feasible) model specifications without detailed contextual guidance. In contrast, the CoT template instructs the LLM to (i) perform a descriptive analysis of the dataset, (ii) apply behavioural constraints commonly used in discrete choice modelling (e.g. negative cost sensitivity), and (iii) generate a structured utility specification using a predefined code syntax (e.g., consistent with Apollo or Biogeme). This strategy simulates common choice modelling workflows \citep{nova2024understanding}.\\

\begin{figure}[H]
  \centering
  \begin{tikzpicture}[
      node distance=1cm and 2cm,
      box/.style={
        draw,
        rectangle,
        minimum width=3cm,
        minimum height=1cm,
        align=center
      },
      >=Stealth
    ]

    \node[box] (prompt) at (-3, 2) {Prompt (ZSP)};
    \node[box] (desc)   at ( 3, 2) {Data Description};

    \node[box] (llm)    at ( 0, 0) {LLM};

    \node[box] (suggest) at ( 0, -2) {Suggest MNL\\Specifications};

    \draw[->] (prompt.south) to[bend right=5] (llm.north);
    \draw[->] (desc.south)   to[bend left=5]  (llm.north);

    \draw[->] (llm.south) -- (suggest.north);

  \end{tikzpicture}
  \caption{Limited Information Setting}
  \label{fig:llm_flowchart2}
\end{figure}
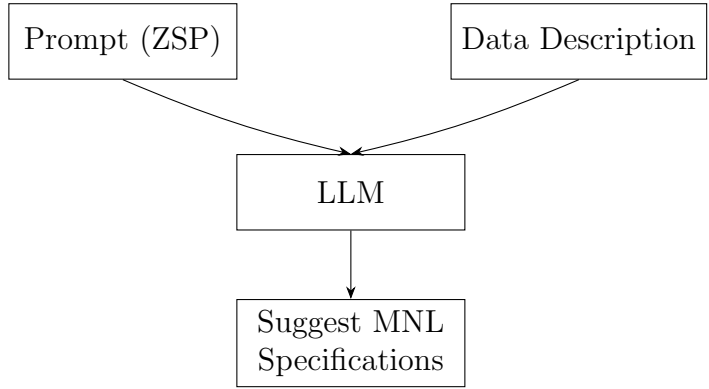

\noindent We use two information settings illustrated in Figures \ref{fig:llm_flowchart1} and \ref{fig:llm_flowchart2}. In the Full Information setting (Figure \ref{fig:llm_flowchart1}), the LLM is provided with a structured input consisting of (i) choice dataset in CSV format, (ii) a structured markdown data description explaining the dataset and its variables, and (iii) a textual prompt (either ZSP or CoT). The LLM is instructed through the provided prompt to either suggest and estimate MNL models to identify the best-fitting MNL specification or to suggest plausible specifications based on the provided data. In the Limited Information setting (Figure \ref{fig:llm_flowchart2}), the LLM is provided only with (i) the data description and (ii) a ZSP prompt strategy. In this setting, the LLM is asked to suggest MNL specifications that are both theoretically sound and expected to perform well empirically, despite not having access to the raw data.\\

\noindent In both modelling goals, ``Suggest" and ``Suggest \& Estimate", LLMs are expected to produce specifications that reflect both standard principles used in discrete choice modelling (e.g. utility specification, expected parameter signs) and the empirical context of the data (e.g. transport or consumer choice). This mirrors a standard modelling workflow, in which a human choice modeller combines methodological knowledge with domain intuition to formulate, refine, and evaluate plausible specifications. Accordingly, the “Suggest” task is designed to assess whether LLMs can generate behaviourally grounded hypotheses in the form of utility specifications that are consistent with established choice modelling principles and likely to perform well empirically. The “Suggest \& Estimate” task extends this assessment to the full modelling workflow, including implementation and estimation, thereby testing whether LLMs can support end-to-end modelling tasks rather than isolated subtasks.\\

\noindent Thus, these experiments vary along three dimensions — \textbf{information setting}, \textbf{prompting strategy}, and \textbf{modelling goal} — resulting in a total of five experimental configurations, as summarized in Table (\ref{tab:Experiment-Setup-Overview}).

\begin{table}[H]
\centering
\caption{Experiment Setup Overview}
\label{tab:Experiment-Setup-Overview}
{%
\begin{tabular}{cccc}
\toprule
Experiment & Information Setting & Prompting Strategy & Modelling Goal \\
\midrule
1 & Full & ZSP & Suggest \& Estimate \\
2 & Full & CoT & Suggest \& Estimate \\
3 & Full & ZSP & Suggest \\
4 & Full & CoT & Suggest \\
5 & Limited & ZSP & Suggest \\
\bottomrule
\end{tabular}%
}
\vspace{0.5em}
\begin{minipage}{\linewidth}
\end{minipage}
\end{table}

\noindent It is important to clarify that the experimental design is not intended to isolate discrete choice modelling knowledge from domain understanding, data interpretation, prompting, or general reasoning. Rather, the objective is to evaluate LLMs as assistive tools in a realistic applied modelling workflow, where these dimensions are naturally combined. The generated utility specifications should therefore be interpreted as integrated outputs reflecting several capabilities, including the interpretation of the choice context, the use of standard behavioural expectations, the selection and transformation of variables, the distinction between generic and alternative-specific parameters, the inclusion of interactions, the need for appropriate identification restrictions, and the construction of estimable model structures. The experiments should therefore be interpreted as a workflow-level assessment of whether current LLMs can generate, and where technically feasible estimate, useful and behaviourally plausible candidate specifications under different prompting and information conditions, rather than as a test of discrete choice modelling knowledge in isolation.

\noindent All MNL specifications generated by the LLMs, whether they are estimated by the LLMs or not, are then implemented and estimated also outside the LLMs. This serves two purposes. First, to verify in the case of ``Suggest \& Estimate'' scenarios that the LLM is estimating the generated specifications correctly and not hallucinating or fabricating the log-likelihood values, as well as the parameter estimates and their significance. Second, for the ``Suggest'' scenarios, all generated specifications are implemented and estimated to assess their empirical performance. Across both goals, all resulting models are evaluated based on goodness-of-fit measures (e.g. log-likelihood, AIC), behavioural plausibility, (e.g. sign and significance of relevant parameters) and expert judgment concerning the overall quality and interpretability of the specifications.

\begin{figure}[H]
  \centering
  \begin{tikzpicture}[
      node distance=1cm and 2cm,
      box/.style={
        draw,
        rectangle,
        minimum width=3.5cm,
        minimum height=1cm,
        align=center
      },
      >=Stealth
    ]

    \node[box] (prompt) at (-4.5, 2) {Prompt (ZSP \textbf{or} CoT)};
    \node[box] (data)   at ( 0, 2) {Data};
    \node[box] (desc)   at ( 4, 2) {Data Description};

    \node[box] (llm)    at ( 0, 0) {LLM};

    \node[box] (combined) at (0, -2) {%
      Suggest \& Estimate\\
      \textbf{or}\\
      Suggest MNL\\
      Specifications%
    };

    \draw[->] (prompt.south) to[bend right=5]   (llm.north);
    \draw[->] (data.south)   --                 (llm.north);
    \draw[->] (desc.south)   to[bend left=5]    (llm.north);

    \draw[->] (llm.south) -- (combined.north);

  \end{tikzpicture}
  \caption{Full Information Setting}
  \label{fig:llm_flowchart1}
\end{figure}
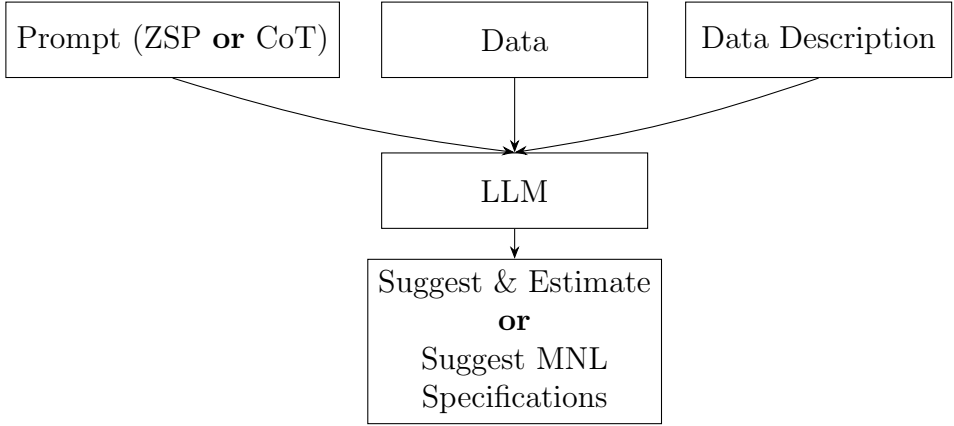

\subsection{Case study setup}

This section presents the different LLMs evaluated, the dataset used, and the prompts we implemented.

\subsubsection{Large Language Models}
We evaluate a diverse set of Large Language Models to investigate how differences in architecture, training scale, openness, and provider influence their ability to support discrete choice model specification. Our selection balances coverage across major proprietary providers (OpenAI, Anthropic, Google, DeepSeek) and includes open-source models (Meta and Google) to ensure transparency and replicability. Moreover, models  were chosen based on their performance, accessibility via browser-based user interfaces or API, and relevance to real-world application scenarios in research and practice. These include GPT-o3 \citep{gpt-o3}, GPT-o4-mini-high \citep{gpt-o3}, and GPT-4o \citep{gpt-45}, Claude 4 Opus \citep{claude4opus}, Claude 4 Sonnet \citep{claude4sonnet}, DeepSeek V3.2 \citep{liu2025deepseek}, Gemini 2.5 Flash \citep{gemini25flash}, Gemma 3 \citep{team2025gemma}, Llama 3 \citep{grattafiori2024llama}, Llama 4 Maverick, Llama 4 Scout \citep{llama4scout}, and Mistral Small (Le Chat) \citep{rastogi2025magistral}.\\

\noindent Table~\ref{tab:llm-capabilities} summarises the capabilities of LLMs evaluated in this study, based on metrics reported by the LLM Leaderboard \citep{llmleaderboard2024}. This comparison provides context on each model's general reasoning abilitiy (e.g., MMLU-Pro, GPQA), coding proficiency (e.g., LiveCodeBench, HumanEval), and performance on human judgement tasks (e.g., HLE) \citep{wang2024mmlu,
rein2024gpqa,
phan2025humanity, tian2024scicode, liang2022holistic}\footnote{The leaderboard compiles and standardizes results from a wide range of benchmark papers, including: arXiv:2501.12948, 2405.04434, 2412.19437, 2412.10302, 2312.11805, 2403.05530, 2303.08774, 2408.12570, 2501.12599, 2410.01257, 2502.00203, 2404.14219, 2412.08905, 2503.01743, 2504.21233, 2504.21318, 2407.10671, 2409.12186, 2309.00071, 2409.12191, 2503.20215, 2502.13923, and 2412.15115.}.

\begin{table}[H]
\centering
\caption{Evaluation metrics for selected LLMs}
\label{tab:llm-capabilities}
\resizebox{\linewidth}{!}{%
\begin{tabular}{l l l l l l l l l l l l l l}
\toprule
Model & Creator & Release & License & Context & MMLU & MMLU-Pro & GPQA & HumanEval & LIVECODEBENCH & SCICODE & HLE \\
\midrule
GPT 4o                  & OpenAI    & 06/08/24 & Proprietary & 128k & 86\% & 75\%  & 46\% & 87\% & 63.5\% & 25\% & 84\% \\
GPT o4-mini             & OpenAI    & 16/04/25 & Proprietary & 200k & 82\% & 65\%  & 40\% & --   & 52.7\% & -& 71\% \\
GPT o3                  & OpenAI    & 16/04/25     & Proprietary & 200k & --   & --    & 83\% & --   &57.1\%    & 33\%    & - \\
Claude 4 Opus           & Anthropic & 22/05/25     & Proprietary & 200k & --   & --    & 83\% & --   & 24.1\%    & -    & 81\% \\
Claude 4 Sonnet         & Anthropic & 22/05/25     & Proprietary & 200k & --   & --    & 84\% & --   & 26.6\%    & -    & 77\% \\
DeepSeek V3.2             & DeepSeek   & 02/12/25 & Open        & 128k & -- & 85\% & 82\% & -- & 83\% & -- & 25\% \\
Gemini 2.5 Flash        & Google    & 20/05/25 & Proprietary & 1m   & --   & 80\%  & 83\% & --   & 50.2\%\% & 36\% & 80\% \\
Llama 3                 & Meta      & 06/12/24 & Open        & 128k & 86\% & 69\%  & 51\% & 88\% & 52.2\%    & 19.8\%    & 74\% \\
Llama 4 Maverick        & Meta      & 05/04/25     & Open        & 1m   & 86\% & 81\%  & 70\% & --   & -    & -    & 79\% \\
Llama 4 Scout           & Meta      & 05/04/25     & Open        & 10m  & 80\% & 74.3\%& 57\% & --   & -    & -    & 76\% \\
Gemma 3 27B                 & Google    & 12/03/25     & Open        & 131k & --   & 68\%  & 42\% & 88\% & -    & -    & 73\% \\
Mistral Small (Le Chat) & Mistral AI & 10/06/25 & Open  & 128k & -- & -- & 68.2\% & -- & 55.8\% & -- & 6.4\% \\
\bottomrule
\end{tabular}%
}
\end{table}

\noindent Interaction with these LLM models was performed mostly via their official browser-based user interfaces, where available (e.g. ChatGPT, Claude, DeepSeek, Gemini). To minimise potential contamination from prior user interactions and ensure greater reproducibility, we took precautions for each LLM. For ChatGPT variants, we disabled memory settings and avoided assigning persistent roles or naming the conversation, ensuring that the model could not accumulate session-specific knowledge. For Claude, Gemini, and DeepSeek, we created new accounts with no prior chat or profile history. These steps were taken to reduce any influence of user-specific context on model outputs. For open-weight models such as Gemma and Llama, interaction was conducted via API calls within local environments to ensure consistency across runs. Moreover, we control by their temperature and probability for suggesting the tokens (see appendix \ref{API}).\\

\noindent Given the inherently probabilistic nature of LLMs, it cannot be guaranteed that repeated queries will produce identical results. However, to improve replicability, each query was executed exactly once per model and per experiment, without regeneration or manual sampling. In addition, we controlled key generation parameters, such as temperature and token probability, when possible (see Appendix~\ref{API}). Our approach prioritises comparability between models under consistent conditions, while recognising that minor variability in results is an expected feature of LLM behaviour.

\subsubsection{Dataset}
\label{dataset}
We evaluate the LLMs using a subset of the inter-city mode choice dataset distributed with the Apollo software \citep{hess2019apollo}. The dataset contains synthetic revealed preference (RP) data from 500 individuals, each reporting  two inter-city trip observations, which yield a total of 1${,}$000 choice responses. The choice set consists of four transport alternatives labelled as car, bus, air, and rail, with at least two of them available per choice situation. Each alternative is characterised by in-vehicle travel time (in minutes), travel cost (in pounds), and access-time (in minutes), the latter only for non-car modes. In addition, the dataset includes information on each individual's gender, income, and whether the journey was a business trip or not.

The synthetic dataset used in this application was originally generated under a Cross-Nested Logit (CNL) structure, while we limit our analysis to the specification and estimation of Multinomial Logit (MNL) specifications. While this may be seen as asking the LLMs to perform model specification and estimation in a mis-specificaton context, two important points need raising.

First, the point of our paper is not to task the LLMs to find the best fitting possible model. The task we set them is to estimate a specific type of model, namely MNL, on the data they are provided with. The fact that the data was generated using a CNL model does not invalidate the results of that process. We can still see which LLMs are able to specify and estimate that model, and which of them obtain the best results, for the task at hand. The true data generating process is in practice never known, and is essentially guaranteed not to be the one used by an analyst, and so our situation actually mimics a real-world scenario (where all models are mis-specified). Second, the argument disappears in any case as our analysis makes use to the RP part of the data only (with the original sample being joint RP-SP), where, for the RP data (which also has a reduced set of attributes), the CNL model collapses to MNL in any case (MNL BIC=2,043.32 \emph{vs} CNL BIC=2,057.33).

\subsubsection{Prompts design}
Prompt design is a critical part of the methodology, as it defines how tasks are framed and used by the LLMs to guide their outputs \citep{white2023prompt}. In this study, prompts were developed through iterative trial-and-error by the authors, drawing on both domain expertise in discrete choice modelling and prior experience interacting with LLMs. While no formal optimisation was conducted, prompts were refined to ensure clarity, internal consistency, and alignment with common modelling workflows \citep{nova2024understanding, van2022choice}. Once established, the final prompt templates were kept fixed across experiments and LLMs to ensure comparability. We acknowledge the growing body of literature on prompt engineering, and agree that systematic prompt development is an important direction for future research \citep{schulhoff2024prompt}. However, this lies beyond the scope of the present study, which focuses on evaluating the modelling capabilities of LLMs under clearly defined and consistently applied prompting strategies. The full prompt texts used for the five different experiment (Table \ref{tab:Experiment-Setup-Overview}) are presented below.

\begin{tcolorbox}[colback=gray!5, colframe=black, fonttitle=\bfseries, title={Experiment 1 
\begin{itemize}[%
    topsep=0pt,
    itemsep=-2pt,
]
    \item Information Setting: Data \& Data Description
    \item Prompting Strategy: Zero-Shot
    \item Modelling Goal: Suggest \& Estimate
    \end{itemize}
    }, enhanced, breakable, sharp corners, boxrule=0.4pt, left=4pt, right=4pt, top=6pt, bottom=6pt]
    You are a choice modeller with over 20 years of experience in discrete choice modelling. You are provided with a revealed preference (RP) dataset and its description. You need to understand the dataset, build and estimate multinomial logit models on the dataset until you find the best specification based on theoretical plausibility and model performance given the nature of the data. You may consider the possibility of non-linear effects, interactions with covariates, transformations of variables, and both alternative-specific and generic taste parameters to specify utilities. Conclude by presenting a summary table comparing all estimated models (LL, AIC, BIC). Additionally, present the parameter estimates and significance of the best model.
\end{tcolorbox}
 
\begin{tcolorbox}[colback=gray!5, colframe=black, fonttitle=\bfseries, title={Experiment 2 
\begin{itemize}[%
    topsep=0pt,
    itemsep=-2pt,
]
    \item Information Setting: Data \& Data Description
    \item Prompting Strategy: Chain-of-Thoughts
    \item Modelling Goal: Suggest \& Estimate
    \end{itemize}
    }, enhanced, breakable, sharp corners, boxrule=0.4pt, left=4pt, right=4pt, top=6pt, bottom=6pt]
    You are a transportation economist with over 20 years of experience in discrete choice modelling. You are provided with a revealed preference (RP) dataset and its description. Your target is to find the best Multinomial Logit (MNL) model specification by relying on this workflow:
\begin{enumerate}
    \item Understand the structure of the dataset and its variables.
    \item Propose, justify, and estimate several utility specifications for a Multinomial Logit (MNL) model until you find the best specification based on theoretical plausibility and model performance in terms of log-likelihood. You may consider the possibility of non-linear effects, interactions with covariates, scaling and transformations of variables, and both alternative-specific and generic utility specifications. Do not split the data into train and test sets.
    \item Recommend the best specification based on theoretical plausibility and model performance.
    \item Present a summary table comparing the specifications using goodness-of-fit metrics (log-likelihood, AIC, BIC, and number of parameters).
    \item Report the parameter estimates and significance for the best-performing model.
\end{enumerate}
\end{tcolorbox}

\begin{tcolorbox}[colback=gray!5, colframe=black, fonttitle=\bfseries, title={Experiment 3 
\begin{itemize}[%
    topsep=0pt,
    itemsep=-2pt,
]
    \item Information Setting: Data \& Data Description
    \item Prompting Strategy: Zero-Shot
    \item Modelling Goal: Suggest
    \end{itemize}
    }, enhanced, breakable, sharp corners, boxrule=0.4pt, left=4pt, right=4pt, top=6pt, bottom=6pt]
    You are a choice modeller with over 20 years of experience in discrete choice modelling. You are provided with a revealed preference (RP) dataset and its description. You need to understand the dataset and propose Multinomial Logit (MNL) model specifications that are both theoretically sound and likely to perform well empirically. The goal is to suggest the best MNL specification based on theoretical plausibility and potential model performance given the nature of the data. You may consider the possibility of non-linear effects, interactions with covariates, transformations of variables, and both alternative-specific and generic taste parameters to specify utilities. Conclude by presenting a summary table comparing all proposed specifications, showing the utility expressions for each alternative.
\end{tcolorbox}

\begin{tcolorbox}[colback=gray!5, colframe=black, fonttitle=\bfseries, title={Experiment 4 
\begin{itemize}[%
    topsep=0pt,      
    itemsep=-2pt,     
]
    \item Information Setting: Data \& Data Description
    \item Prompting Strategy: Chain-of-Thoughts
    \item Modelling Goal: Suggest
    \end{itemize}
    }, enhanced, breakable, sharp corners, boxrule=0.4pt, left=4pt, right=4pt, top=6pt, bottom=6pt]
    You are a transportation economist with over 20 years of experience in discrete choice modelling. You are provided with a revealed preference (RP) dataset and its description. Your target is to find the best Multinomial Logit (MNL) model specification by relying on this workflow:
\begin{enumerate}
    \item Understand the structure of the dataset and its variables.
    
    \item Perform descriptive analysis and assess the distribution of both continuous and categorical variables. 
    
    \item Evaluate alternative shares (choice frequencies) to understand their market shares and potential data imbalance.
    
    \item Conduct a correlation analysis among independent variables to identify potential collinearity. Discuss how collinearity issues, if identified, should be managed.
    
    \item Conduct cross-tabulations between covariates and chosen alternatives to understand behavioral patterns and segmentation clearly.
    
    \item Compute and interpret average values of continuous variables grouped by chosen alternatives.
    
    \item Based on your descriptive analysis and understanding of the data, propose and justify several utility specifications for a Multinomial Logit (MNL) model that would fit the data best based on theoretical plausibility and potential model performance. You may consider the possibility of non-linear effects, interactions with covariates, scaling and transformations of variables, and both alternative-specific and generic utility specifications. Do not split the data into train and test sets.

    \item Conclude by presenting a summary table comparing all proposed specifications, showing the utility expressions for each alternative.

\end{enumerate}
\end{tcolorbox}

\begin{tcolorbox}[colback=gray!5, colframe=black, fonttitle=\bfseries, title={Experiment 5 
\begin{itemize}[%
    topsep=0pt,
    itemsep=-2pt,
]
    \item Information Setting: Data Description
    \item Prompting Strategy: Zero-Shot
    \item Modelling Goal: Suggest
    \end{itemize}
    }, enhanced, breakable, sharp corners, boxrule=0.4pt, left=4pt, right=4pt, top=6pt, bottom=6pt]
    You are a choice modeller with over 20 years of experience in discrete choice modelling. You are provided a description of a dataset that contains revealed preference (RP) observations. You need to understand the dataset and propose Multinomial Logit (MNL) model specifications that are both theoretically sound and likely to perform well empirically. The goal is to suggest the best MNL specification based on theoretical plausibility and potential model performance given the nature of the data. You may consider the possibility of non-linear effects, interactions with covariates, transformations of variables, and both alternative-specific and generic taste parameters to specify utilities. Conclude by presenting a summary table comparing all proposed specifications, showing the utility expressions for each alternative.
\end{tcolorbox}

\section{Results}
\label{section:res}
This section presents the empirical results. Section \ref{results_1} provides a detailed evaluation of the performance of the different LLMs in each experimental configuration, with particular attention to how prompting strategies and information availability influence the quality of suggested model specifications. Section \ref{results_2} offers an overall evaluation of the utility specifications generated by each LLM, independent of experimental setting.\\

\noindent As discussed in Section \ref{dataset}, the mode choice dataset used in this application was originally generated under a Cross-Nested Logit structure on joint RP-SP data, while the prompting strategies in this study restrict the LLMs to propose MNL specifications only on the RP subset (which includes a subset of the attributes). As a result, a direct comparison with the true data-generating parameters are not possible. Instead, the proposed specifications are evaluated using (i) goodness-of fit metrics and (ii) behavioural expectations in terms of parameter signs.

\subsection{Evaluation by Experiment}
\label{results_1}
\subsubsection{Goal: Suggest \& Estimate}
We start by presenting the outcomes of the two ``Suggest \& Estimate" experiments, which differ in their prompting strategies. The first experiment is executed with a zero-shot prompt, while the second experiment applies a chain-of-thought prompt (Table \ref{tab:Experiment-Setup-Overview}). Among all evaluated LLMs, only ChatGPT-o3 successfully suggested valid MNL specifications and produced estimation results by executing self-generated Python code that could be independently replicated.\footnote{See ChatGPT o3 transcript of Experiment 1: https://chatgpt.com/share/6887d3a6-815c-8009-9b40-78be528ef4f7}. Specifically, all specifications generated and estimated by ChatGPT-o3 were re-implemented and estimated in Apollo \citep{hess2019apollo}. The resulting log-likelihood values and parameter estimates matched those reported by ChatGPT-o3 up to a numerical tolerance, with relative discrepancies below 1\% for a small number of values. In contrast, every other LLM either mis-estimated specifications it suggested or hallucinated the results by returning log-likelihood values, parameter estimates and standard deviations that we could not reproduce in our independent re-estimations. This finding should not be interpreted as a general limitation of LLMs' reasoning capabilities, but rather as a consequence of architectural and interface differences across the evaluated models. In our experimental setup, ChatGPT-o3 operated in an agentic mode with access to external computational tools, most importantly, to load the attached dataset as a file and execute self-generated Python code in an execution environment. These tools enabled ``correct" end-to-end estimation. By contrast, the other LLMs we tested were non-agentic in our setup and therefore lacked execution environments. As a result, they lacked the ability to execute self-generated code, if any, which contributed to mis-estimation and/or hallucinated outputs. This places a responsibility on modellers to account for architectural differences across LLMs when delegating modelling tasks and to independently verify any estimation results produced by such systems.\\

\noindent Tables \ref{tab:exp1} and \ref{tab:exp2} summarise the results of the specifications suggested and estimated by ChatGPT o3 across the first and second experiments. The zero-shot prompt generated and estimated three specifications, while the chain-of-thought prompt produced five. Despite the difference in prompt, both experiments converged to the same best-performing specification in terms of Log-Likelihood (LL), Akaike Information Criterion (AIC), and Bayesian Information Criterion (BIC) (S2 in Table \ref{tab:exp1} and S4 in Table \ref{tab:exp2}). The utility functions of the best-performing model follows a linear-in-parameter specification. They include generic taste parameters for travel time, travel cost, and access time (the latter for all modes except car), as well as a generic interaction coefficient between travel time and a business trip variable. The specification also incorporates alternative-specific constants, with the constant for the car alternative normalised to zero for identification.

\begin{table}[H]
\centering
\caption{Experiment 1 (Full Information, ZSP, Suggest \& Estimate)}
\label{tab:exp1}
{%
\begin{tabular}{c c c c c}
\toprule
\makecell{Specification\\(ChatGPT o3)} & LL & AIC & BIC & VOT \\
\midrule
S1 & $-1{,}031.82$ & $2{,}073.63$ & $2{,}098.17$ & $0.197$ \\
$\mathbf{S2}$ & $\mathbf{-981.80}$ & $\mathbf{1{,}977.61}$ & $\mathbf{2{,}011.96}$ & $0.198$ \\
S3 & $-1{,}083.68$ & $2{,}177.36$ & $2{,}201.90$ & $0.262$ \\
\bottomrule
\end{tabular}%
}
\end{table}

\begin{table}[H]
\centering
\caption{Experiment 2 (Full Information, CoT, Suggest \& Estimate)}
\label{tab:exp2}
{%
\begin{tabular}{c c c c c}
\toprule
\makecell{Specification\\(ChatGPT o3)} & LL & AIC & BIC & VOT \\
\midrule
S1 & $-1{,}031.82$ & $2{,}073.63$ & $2{,}098.17$ & $0.197$ \\
S2 & $-1{,}030.97$ & $2{,}073.93$ & $2{,}103.38$ & $0.198$ \\
S3 & $-1{,}026.81$ & $2{,}065.62$ & $2{,}095.06$ & $0.262$ \\
$\mathbf{S4}$ & $\mathbf{-981.80}$ & $\mathbf{1{,}977.61}$ & $\mathbf{2{,}011.96}$ & $0.198$ \\
S5 & $-993.30$ & $2{,}000.59$ & $2{,}034.95$ & $0.364$ \\
\bottomrule
\end{tabular}%
}
\end{table}

\subsubsection{Goal: Suggest}
Next, we discuss the results of the ``Suggest'' experiments, which explore how different prompting strategies and information availability influence the quality of model specifications generated by LLMs. Experiments 3 and 4 are both under the same full information setup (providing the model with both the raw dataset and a structured
data description) but differ in prompting strategies (ZSP for Experiment 3 and CoT for Experiment 4). In contrast, Experiment 5 is conducted with limited information, providing only the data description, and a ZSP (Table \ref{tab:Experiment-Setup-Overview}). Note that the Llama models (Llama 3, Llama 4 Maverick, and Llama 4 Scout) were not included in Experiments 3 and 4, as their API interface did not support loading raw data files at the time of evaluation. As a result, these models were only evaluated under the limited-information setting in Experiment 5.\\

\noindent The results from Experiment 3 demonstrate that LLMs, when prompted under a ZSP strategy and provided with full information, are capable of generating plausible and well-performing model specifications; though not without limitations. Tables \ref{tab:exp3-ll} and \ref{tab:exp3-aic} present the LL and AIC values, respectively, for all specifications generated by nine LLM variants in Experiment 3. Notably, the Claude family (Claude 4 Opus and 4 Sonnet) and Mistral le Chat produced a greater number of specifications than any ChatGPT or Gemini variant, which reflects a slightly higher capacity for exploring the modelling space. All specifications suggested by ChatGPT 4o systematically omitted alternative-specific constants, which are important in the case of labelled alternatives. As a result, these specifications obtained far inferior fit and were excluded from further comparison. In addition, two specifications generated by Claude 4 Opus failed to converge (S2 and S5) and one specification from Gemini 2.5 Flash (S4) yielded at least one positive coefficient for travel time and/or travel, which violates economic behaviour expectations. These were also excluded. All  excluded specifications are shown in gray font within the tables for transparency. In contrast, all specifications generated by ChatGPT o4-mini-high, ChatGPT o3, Claude 4 Sonnet, Mistral le Chat, DeepSeek V3.2, and Gemma 3 passed both convergence and behavioural plausibility checks. Among the valid specifications, specification 5 (S5) from DeepSeek V3.2 has the best LL ($\text{LL} = -970.47$) and AIC ($\text{AIC} = 1{,}960.93$), demonstrating a more favourable balance between model complexity and fit.\\

\noindent Overall, the valid specifications generated in the ``Suggest'' experiment exhibit slightly improved goodness-of-fit relative to those identified in the ``Suggest \& Estimate'' configurations. This suggests that, when LLMs are tasked solely with proposing specifications, without the additional workload involved in estimating them, they may provide models with marginal better goodness-of-fit metrics.  One possible explanation is that separating specification task from estimation enables the LLMs to fully allocate their computational resources and internal reasoning capabilities to generating behaviourally plausible and empirically well-performing  model specifications. In contrast, executing the estimation process introduces  substantial computation demands, including data handling, code generation, syntax validation, and numerical optimisation, all of which may detract from the LLM’s ability to focus on the specification task itself.

\begin{table}[H]
\centering
\caption{Experiment 3 (Full Information, ZSP, Suggest) - LL}
\label{tab:exp3-ll}
\resizebox{\linewidth}{!}{%
\begin{tabular}{c c c c c c c c c c c}
\toprule
Spec. & \makecell{ChatGPT\\4o} & \makecell{ChatGPT\\o4-mini-high} & \makecell{ChatGPT\\o3} & \makecell{Claude\\4 Opus} & \makecell{Claude\\4 Sonnet} & \makecell{Mistral\\le Chat} & \makecell{DeepSeek\\V3.2} & \makecell{Gemini\\2.5 Flash} & \makecell{Gemma\\3}\\
\midrule
S1 & $\textcolor{gray}{-1{,}106.23}^*$ & $-1{,}031.82$ & $-1{,}031.00$ & $-1{,}030.97$ & $-1{,}031.00$ & $-1{,}031.82$ & $-1{,}031.00$ & $-1{,}031.82$ & $-1{,}030.97$ \\
S2 & $\textcolor{gray}{-1{,}027.57}^*$ & $-1{,}030.97$ & $-1{,}030.97$ & $\textcolor{gray}{-1{,}048.30}^{\dagger}$ & $-1{,}025.91$ & $-1030.97$ & $-1{,}030.97$ & $-1{,}030.97$ & $-999.70$ \\
S3 & $\textcolor{gray}{-1{,}022.32}^*$ & $-1{,}025.00$ & $-1{,}036.49$ & $-1{,}026.10$ & $-1{,}017.58$ & $-1{,}022.65$ & $-1{,}027.62$ & $\mathbf{-983.45}$ & $-1{,}000.05$ \\
S4 & $\textcolor{gray}{-1{,}023.33}^*$ & $\mathbf{-999.72}$ & $-1{,}011.22$ & $\mathbf{-982.15}$ & $\mathbf{-978.37}$ & $-1{,}035.53$ & $-1{,}021.96$ & $\textcolor{gray}{-1{,}028.78}^{\ddagger}$ & $\mathbf{-998.60}$\\
S5 & $\textcolor{gray}{-1{,}123.26}^*$ & $-1{,}036.38$ & $\mathbf{-977.37}$ & $\textcolor{gray}{-1{,}033.34}^{\dagger}$ & $\mathbf{-978.37}$ & $\mathbf{-982.69}$ & $\mathbf{-970.47}$ & -- & --\\
S6 & -- & -- & -- & $-1{,}027.99$ & $-1{,}030.78$ & $-1{,}031.78$ & -- & -- & -- \\
\bottomrule
\end{tabular}%
}
\vspace{-2ex}
  {\scriptsize
    \begin{flushleft}
      $^*$No ASCs included. $^{\dagger}$Model did not converge. $^{\ddagger}$Positive Beta Cost and/or Beta Time.
    \end{flushleft}
  }
\end{table}

\begin{table}[H]
\centering
\caption{Experiment 3 (Full Information, ZSP, Suggest) - AIC}
\label{tab:exp3-aic}
\resizebox{\linewidth}{!}{%
\begin{tabular}{c c c c c c c c c c c}
\toprule
Spec. & \makecell{ChatGPT\\4o} & \makecell{ChatGPT\\o4-mini-high} & \makecell{ChatGPT\\o3} & \makecell{Claude\\4 Opus} & \makecell{Claude\\4 Sonnet} & \makecell{Mistral\\le Chat} & \makecell{DeepSeek\\V3.2} & \makecell{Gemini\\2.5 Flash} & \makecell{Gemma\\3}\\
\midrule
S1 & $\textcolor{gray}{2{,}216.47}^*$ & $2{,}073.63$ & $2{,}071.99$ & $2{,}073.93$ & $2{,}071.99$ & $2{,}073.63$ & $2{,}071.99$ & $2{,}073.63$ & $2{,}073.93$\\
S2 & $\textcolor{gray}{2{,}071.14}^*$ & $2{,}073.94$ & $2{,}073.93$ & $\textcolor{gray}{2{,}108.60}^{\dagger}$ & $2{,}065.82$ & $2{,}073.94$ & $2{,}073.93$ & $2{,}073.93$ & $\mathbf{2{,}033.39}$\\
S3 & $\textcolor{gray}{2{,}060.64}^*$ & $2{,}074.01$ & $2{,}084.98$ & $2{,}070.19$ & $2{,}053.15$ & $2{,}057.31$ & $2{,}079.24$ & $\mathbf{1{,}986.90}$ & $2{,}034.10$\\
S4 & $\textcolor{gray}{2{,}062.67}^*$ & $\mathbf{2{,}029.44}$ & $2{,}036.44$ & $\mathbf{1{,}980.31}$ & $\mathbf{1{,}976.75}$ & $2{,}081.07$ & $2{,}057.92$ & $\textcolor{gray}{2{,}073.56}^{\ddagger}$ & $2{,}037.19$ \\
S5 & $\textcolor{gray}{2{,}250.53}^*$ & $2{,}084.77$ & $\mathbf{1{,}984.74}$ & $\textcolor{gray}{2{,}090.67}^{\dagger}$ & $\mathbf{1{,}976.75}$ & $\mathbf{1{,}977.39}$ & $\mathbf{1{,}960.93}$ & -- & -- \\
S6 & -- & -- & -- & $2{,}067.98$ & $2{,}073.56$ & $2{,}075.56$ &  & -- & -- \\
\bottomrule
\end{tabular}%
}
\vspace{-2ex}
  {\scriptsize
    \begin{flushleft}
      $^*$No ASCs included. $^{\dagger}$Model did not converge. $^{\ddagger}$Positive Beta Cost and/or Beta Time.
    \end{flushleft}
  }
\end{table}

\noindent Next, we present the LL and AIC of all specifications generated in Experiment 4 (Tables \ref{tab:exp4-ll} and \ref{tab:exp4-aic}). Unlike the zero-shot prompt in Experiment 3, this experiment employed a structured chain-of-thought prompting strategy, encouraging the LLMs to perform preliminary analyses before suggesting utility specifications. The results of Experiment 4 demonstrate that structured prompting, in the form of chain-of-thought strategies, can substantially improve the quality of LLM-suggested model specifications. By guiding LLM models through intermediate analysis steps before suggesting model specifications, CoT prompting encourages more behaviourally plausible and complex utility forms. \\

\noindent As shown in Tables \ref{tab:exp4-ll} and \ref{tab:exp4-aic}, this approach led to consistent improvements across various aspects compared to the ZSP baseline used in Experiment 3. Most notably, the use of CoT resolved a recurring issue from Experiment 3 in which alternative-specific constants were sometimes omitted; now, all generated specifications included them. Furthermore, CoT prompting improved the overall model fit across most LLMs. Specifically, all LLMs, except Mistral le Chat and DeepSeek V3.2, achieved better AIC values under CoT prompting (Experiment 4, Table \ref{tab:exp4-aic}) than under the ZSP condition (Experiment 3, Table \ref{tab:exp3-aic}). Moreover, ChatGPT 4o, ChatGPT o4-mini-high, Claude 4 Opus and 4 Sonnet, and Gemini 2.5 suggested better specifications in terms of LL values. Despite these gains, certain limitations persisted. One specification from DeepSeek V3.2 and two specifications from Mistral le Chat failed to converge during estimation. In addition, one specification from ChatGPT 4o and two specifications from Gemini 2.5 Flash led to at least one positive coefficient for travel time and/or travel cost, contradicting expected economic behaviour. Surprisingly, all specifications generated by the open-weight Gemma 3 from Google Deepmind failed to converge during estimation. Despite these limitations, the best-performing model overall, Specification 5 from Claude 4 Sonnet, achieved the highest log-likelihood ($-967.53$) and the lowest AIC ($1{,}963.05$), demonstrating the potential of CoT prompts to guide LLMs toward more effective and interpretable model structures.

\begin{table}[H]
\centering
\caption{Experiment 4 (Full Information, CoT, Suggest) - LL}
\label{tab:exp4-ll}
\resizebox{\linewidth}{!}{%
\begin{tabular}{c c c c c c c c c c c}
\toprule
Spec. & \makecell{ChatGPT\\4o} & \makecell{ChatGPT\\o4-mini-high} & \makecell{ChatGPT\\o3} & \makecell{Claude\\4 Opus} & \makecell{Claude\\4 Sonnet} & \makecell{Mistral\\le Chat} & \makecell{DeepSeek\\V3.2} & \makecell{Gemini\\2.5 Flash} & \makecell{Gemma\\3}\\
\midrule
S1 & $-1{,}024.48$ & $-1{,}030.97$ & $-1{,}030.97$ & $-1{,}024.48$ & $-1{,}019.18$ & $\mathbf{-1{,}030.97}$ & $-1{,}032.28$ & $-1{,}031.42$ & \textcolor{gray}{$-1{,}030.97$}$^{\dagger}$ \\
S2 & $\mathbf{-979.47}$ & $-1{,}024.48$ & $\mathbf{-980.62}$ & $-1{,}030.97$ & $-1{,}010.30$ & \textcolor{gray}{$-1{,}030.97$}$^{\dagger}$ & $-1{,}025.01$ & \textcolor{gray}{$-1{,}024.36$}$^{\ddagger}$ & \textcolor{gray}{$-1{,}030.97$}$^{\dagger}$\\
S3 & $-1{,}024.97$ & $\mathbf{-987.79}$ & $-1{,}026.00$ & $\mathbf{-973.34}$ & $-1{,}019.76$ & $-1{,}036.49$ & $-1{,}021.96$ & $\mathbf{-968.29}$ & \textcolor{gray}{$-1{,}024.84$}$^{\dagger}$\\
S4 & \textcolor{gray}{$-1{,}017.16$}$^{\ddagger}$ & $-1{,}036.49$ & $-1{,}035.45$ & $-1{,}035.14$ & $-1{,}024.00$ & \textcolor{gray}{$-1{,}030.97$}$^{\dagger}$ & $-1{,}027.52$ & $-1{,}024.48$ & \textcolor{gray}{$-1{,}028.78$}$^{\dagger}$\\
S5 & -- & $-1{,}014.87$ & -- & $-1{,}040.96$ & $\mathbf{-967.53}$ & -- & $\mathbf{-980.95}$ & $-1{,}030.97$ & -- \\
S6 & -- & -- & -- & $-980.82$ & $-1{,}019.32$ & -- & \textcolor{gray}{$-979.64$}$^{\dagger}$ & \textcolor{gray}{$-960.07$}$^{\ddagger}$ & -- \\
\bottomrule
\end{tabular}%
}
\vspace{-2ex}
  {\scriptsize
    \begin{flushleft}
      $^{\dagger}$Model did not converge. $^{\ddagger}$Positive Beta Cost and/or Beta Time.
    \end{flushleft}
  }
\end{table}

\begin{table}[H]
\centering
\caption{Experiment 4 (Full Information, CoT, Suggest) - AIC}
\label{tab:exp4-aic}
\resizebox{\linewidth}{!}{%
\begin{tabular}{c c c c c c c c c c c}
\toprule
Spec. & \makecell{ChatGPT\\4o} & \makecell{ChatGPT\\o4-mini-high} & \makecell{ChatGPT\\o3} & \makecell{Claude\\4 Opus} & \makecell{Claude\\4 Sonnet} & \makecell{Mistral\\le Chat} & \makecell{DeepSeek\\V3.2} & \makecell{Gemini\\2.5 Flash} & \makecell{Gemma\\3} \\
\midrule
S1 & $2{,}076.96$ & $2{,}073.93$ & $2{,}073.94$ & $2{,}076.96$ & $2{,}056.37$ & $\mathbf{2{,}079.93}$ & $2{,}076.56$ & $2{,}074.84$ & \textcolor{gray}{$2{,}079.93$}$^{\dagger}$ \\
S2 & $\mathbf{1{,}992.94}$ & $2{,}076.95$ & $\mathbf{1{,}979.23}$ & $2{,}073.93$ & $2{,}038.61$ & \textcolor{gray}{$2{,}075.93$}$^{\dagger}$ & $2{,}074.01$ & \textcolor{gray}{$2{,}080.72$}$^{\ddagger}$ & \textcolor{gray}{$2{,}081.93$}$^{\dagger}$\\
S3 & $2{,}077.94$ & $\mathbf{1{,}993.59}$ & $2{,}070.01$ & $\mathbf{1{,}970.69}$ & $2{,}055.52$ & $2{,}084.92$ & $2{,}057.93$ & $\mathbf{1{,}966.59}$ & \textcolor{gray}{$2{,}067.67$}$^{\dagger}$\\
S4 & \textcolor{gray}{$2{,}068.32$}$^{\ddagger}$ & $2{,}084.98$ & $2{,}082.89$ & $2{,}082.28$ & $2{,}066.01$ & \textcolor{gray}{$2{,}077.93$}$^{\dagger}$ & $2{,}067.05$ & $2{,}076.95$ & \textcolor{gray}{$2{,}077.56$}$^{\dagger}$\\
S5 & -- & $2{,}043.74$ & -- & $2{,}091.92$ & $\mathbf{1{,}963.05}$ & -- & $\mathbf{1{,}977.90}$ & $2{,}079.93$ & -- \\
S6 & -- & -- & -- & $1{,}977.63$ & $2{,}052.64$ & -- & \textcolor{gray}{$1{,}987.28$}$^{\dagger}$ & \textcolor{gray}{$1{,}972.14$}$^{\ddagger}$ & -- \\
\bottomrule
\end{tabular}%
}
\vspace{-2ex}
  {\scriptsize
    \begin{flushleft}
      $^{\dagger}$Model did not converge. $^{\ddagger}$Positive Beta Cost and/or Beta Time.
    \end{flushleft}
  }
\end{table}

\noindent The results of Experiment 5 indicate that LLMs are capable of generating well-performing model specifications even under limited information conditions, where only a dataset description is provided. Despite the absence of raw data, several LLMs produced specifications that equalled or even surpassed those they generated under full information conditions in previous experiments. As shown in Tables \ref{tab:exp5-ll} and \ref{tab:exp5-aic}, Claude variants and Mistral le Chat continued to show stronger generative capacity, producing a greater number of  specifications than any of the GPT or Gemini models. One specification from GPT-o4-mini-high was omitted for lacking ASCs, which led to poor model fit, while  one specification from Llama 4 Scout was excluded due to convergence failures. Moreover, none of the specifications generated by the open-weight Gemma 3 model converged during estimation and were therefore excluded from further analysis. Among the valid results, Specification 5 (S5) from DeepSeek V3.2 achieved the best log-likelihood ($-956.68$) and best AIC ($1{,}945.36$), indicating a favourable trade-off between model fit and parsimony. Notably, ChatGPT 4o, o4-mini-high, Claude 4 Opus, Mistral le Chat, and DeepSeek all produced better-fitting specifications in this limited information setting than in their respective full information experiments (Experiments 3 and 4). These findings suggest that, under certain conditions, LLMs may benefit from simplified input structures, enabling them to focus more effectively on the model specification process itself. \\

\noindent Finally, open-weight models (Gemma 3 and Llama variants) showed noticeably weaker performance in Experiment 5. While the limited information setup reduced the task complexity, these models struggled to deliver viable or competing specifications. As previously mentioned, none of the specifications generated by Gemma 3 converged. The Llama family performed somewhat better, with Llama 4 Scout generating a specification (S5) with a moderately competitive LL (-981.98) and AIC (1,997.96), though still weaker than most specifications generated by closed-weight models. However, the best specifications from the other two Llama versions (3 and 4 Maverick) produced the worst best specifications in terms of LL and AIC compared to all closed-weight models. These findings confirm that current open-weight LLMs, while valuable for transparency and local deployment, are still far from matching the end-to-end reliability and modeling capacity of closed-weight models in discrete choice modeling tasks, even under simplified tasks with limited information conditions.

\begin{table}[H]
\centering
\caption{Experiment 5 (Limited Information, ZSP, Suggest) - LL}
\label{tab:exp5-ll}
\resizebox{\linewidth}{!}{%
\begin{tabular}{c c c c c c c c c c c}
\toprule
Spec. & \makecell{ChatGPT\\4o} & \makecell{ChatGPT\\o4-mini-high} & \makecell{ChatGPT\\o3} & \makecell{Claude\\4 Opus} & \makecell{Claude\\4 Sonnet} & \makecell{Mistral\\le Chat} & \makecell{DeepSeek\\V3.2} & \makecell{Gemini\\2.5 Flash} & \makecell{Gemma\\3} \\
\midrule
S1 & $-1{,}031.00$ & $\textcolor{gray}{-1{,}121.16}^*$ & $-1{,}031.00$ & $-1{,}031.82$ & $-1{,}030.97$ & $-1{,}031.00$ & $-1{,}031.82$ & $-1{,}031.82$ & \textcolor{gray}{$-1{,}024.48$}$^{\dagger}$ \\
S2 & $\mathbf{-969.73}$ & $-1{,}030.97$ & $-1{,}030.97$ & $-1{,}030.97$ & $-1{,}026.81$ & $-1{,}025.64$ & $-1{,}026.34$ & $-1{,}030.97$ & \textcolor{gray}{$-1{,}022.34$}$^{\dagger}$ \\
S3 & $-976.60$ & $-1{,}036.47$ & $-1{,}015.31$ & $-1{,}022.61$ & $-1{,}023.22$ & $-1{,}035.77$ & $-1{,}025.00$ & $\mathbf{-981.57}$ & \textcolor{gray}{$-960.90$}$^{\dagger}$ \\
S4 & -- & $\mathbf{-981.47}$ & $\mathbf{-982.60}$ & $\mathbf{-972.40}$ & $\mathbf{-971.33}$ & $-1{,}021.97$ & $-958.47$ & $-1{,}024.48$ & \textcolor{gray}{$-955.06$}$^{\dagger}$\\
S5 & -- & $-1{,}020.82$ & $-1{,}014.85$ & $-1{,}035.14$ & $-978.89$ & $-1{,}025.30$ & $\mathbf{-956.68}$ & -- & \textcolor{gray}{$-955.18$}$^{\ddagger}$ \\
S6 & -- & -- & -- & $-1{,}030.26$ & $-972.07$ & $-1{,}030.28$ & -- & -- & -- \\
S7 & -- & -- & -- & $-1{,}006.36$ & -- & $\mathbf{-974.06}$ & -- & -- & -- \\
\toprule
Spec. & \makecell{Llama\\3} & \makecell{Llama\\4 Maverick} & \makecell{Llama\\4 Scout} & & & & & & & \\
\midrule
S1 & $-1{,}024.48$ & $-1{,}030.97$ & $-1{,}031.82$ & & & & & & & \\
S2 & $\mathbf{-991.85}$ & $-1{,}024.48$ & $-1{,}012.41$ & & & & & & & \\
S3 & $-1{,}024.44$ & $\mathbf{-1{,}021.96}$ & $-1{,}012.41$ & & & & & & & \\
S4 & $-1{,}000.18$ & -- & \textcolor{gray}{$-1{,}010.09$}$^{\dagger}$ & & & & & & & \\
S5 & -- & -- & $\mathbf{-981.98}$ & & & & & & & \\
\bottomrule
\end{tabular}%
}
\vspace{-2ex}
  {\scriptsize
    \begin{flushleft}
      $^*$No ASCs included. $^{\dagger}$Model did not converge.
      $^{\ddagger}$Positive Beta Cost and/or Beta Time.
    \end{flushleft}
  }
\end{table}

\begin{table}[H]
\centering
\caption{Experiment 5 (Limited Information, ZSP, Suggest) - AIC}
\label{tab:exp5-aic}
\resizebox{\linewidth}{!}{%
\begin{tabular}{c c c c c c c c c c c}
\toprule
Spec. & \makecell{ChatGPT\\4o} & \makecell{ChatGPT\\o4-mini-high} & \makecell{ChatGPT\\o3} & \makecell{Claude\\4 Opus} & \makecell{Claude\\4 Sonnet} & \makecell{Mistral\\le Chat} & \makecell{DeepSeek\\V3.2} & \makecell{Gemini\\2.5 Flash} & \makecell{Gemma\\3} \\
\midrule
S1 & $2{,}071.99$ & $\textcolor{gray}{2{,}246.32}^*$ & $2{,}071.99$ & $2{,}073.63$ & $2{,}073.93$ & $2{,}071.99$ & $2{,}073.63$ & $2{,}073.63$ & \textcolor{gray}{$2{,}078.95$}$^{\dagger}$ \\
S2 & $\mathbf{1{,}973.46}$ & $2{,}073.93$ & $2{,}073.93$ & $2{,}073.93$ & $2{,}065.62$ & $2{,}073.28$ & $2{,}074.67$ & $2{,}073.93$ & \textcolor{gray}{$2{,}090.68$}$^{\dagger}$ \\
S3 & $1{,}987.19$ & $2{,}084.93$ & $2{,}046.62$ & $2{,}059.21$ & $2{,}064.46$ & $2{,}081.54$ & $2{,}074.01$ & $\mathbf{1{,}983.14}$ & \textcolor{gray}{$1{,}969.80$}$^{\dagger}$ \\
S4 & -- & $\mathbf{1{,}980.95}$ & $\mathbf{1{,}991.20}$ & $\mathbf{1{,}960.81}$ & $1{,}972.65$ & $2{,}055.95$ & $1{,}946.95$ & $2{,}076.96$ & \textcolor{gray}{$1{,}974.12$}$^{\dagger}$ \\
S5 & -- & $2{,}057.65$ & $2{,}047.70$ & $2{,}082.28$ & $1{,}987.77$ & $2{,}064.60$ & $\mathbf{1{,}945.36}$ & -- & \textcolor{gray}{$1{,}986.36$}$^{\ddagger}$ \\
S6 & -- & -- & -- & $2{,}078.52$ & $\mathbf{1{,}972.15}$ &$2{,}076.57$ & -- & -- & -- \\
S7 & -- & -- & -- & $2{,}028.72$ & -- & $\mathbf{1{,}970.13}$ & -- & -- & -- \\
\toprule
Spec. & \makecell{Llama\\3} & \makecell{Llama\\4 Maverick} & \makecell{Llama\\4 Scout} & & & & & & & \\
\midrule
S1 & $2{,}076.95$ & $2{,}073.97$ & $2{,}073.63$ & & & & & & & \\
S2 & $\mathbf{2{,}021.70}$ & $2{,}076.95$ & $2{,}058.82$ & & & & & & & \\
S3 & $2{,}076.88$ & $\mathbf{2{,}057.92}$ & $2{,}058.82$ & & & & & & & \\
S4 & $2{,}036.36$ & -- & \textcolor{gray}{$2{,}070.18$}$^{\dagger}$ & & & & & & & \\
S5 & -- & -- & $\mathbf{1{,}997.96}$ & & & & & & & \\
\bottomrule
\end{tabular}%
}
\vspace{-2ex}
  {\scriptsize
    \begin{flushleft}
      $^*$No ASCs included. $^{\dagger}$Model did not converge.
      $^{\ddagger}$Positive Beta Cost and/or Beta Time.
    \end{flushleft}
  }
\end{table}

\subsubsection{Best Specifications by Experiment and/or LLM}

To evaluate the capabilities of each Large Language Model across experimental conditions, we analyse the best-performing model specifications in terms of LL, AIC, and BIC. This analysis offers an overview of how LLMs perform when tasked with either suggesting or estimating MNL models under varying information and prompt settings. Tables~\ref{tab:best-ll-exp}, ~\ref{tab:best-aic-exp}, and~\ref{tab:best-bic-exp} summarise the best specifications by experiment and by LLM.\\

\noindent  In the ``Suggest \& Estimate'' experiments (Exp. 1 and 2), only GPT-o3 successfully specified theoretically sound specifications and estimated them correctly through self-generated Python code, regardless of the prompting strategy. All other LLMs either mis-estimated their own specifications or returned fabricated estimation outputs. In contrast, when LLMs were tasked solely with specification (Exp. 3 to 5), the competitive performance among LLMs improved in terms of LL, AIC, and BIC. Under the full information and ZSP condition (Exp. 3), DeepSeek V3.2 generated the specification with the best log-likelihood ($\text{LL} = -970.47$). With a structured CoT prompt in Experiment 4, Claude 4 Sonnet produced the best specification in terms of log-likelihood ($\text{LL} = -967.53$). Lastly, in the limited information setting (Exp. 5), DeepSeek V3.2 produced not only the best specification of that experiment but the best specification across all experiments, achieving a log-likelihood of $-956.68$. Surprisingly, five LLMs (GPT 4o, GPT o4-mini-high, Claude 4 Opus, Mistral le Chat, and DeepSeek) achieved their best performance under this limited-input configuration, suggesting that withholding full data access may, in some cases, prompt stronger theoretical reasoning.\\

\noindent Similarly to the LL ranking, DeepSeek V3.2 provided the best AIC values in Experiments 3 and 5, whereas Claude 4 Sonnet yielded the best AIC in Experiment 4 (Table ~\ref{tab:best-aic-exp}). Moreover, DeepSeek V3.2, under limited information conditions (Exp. 5), generated the specification with the overall lowest AIC ($1{,}945.36$). In line with the LL results, the same five LLMs (GPT 4o, GPT o4-mini-high, Claude 4 Opus, Mistral le Chat, and DeepSeek) achieved their respective best performances in the limited information experiment.\\

\begin{table}[H]
\centering
\caption{Experiments Summary - LL}
\label{tab:best-ll-exp}
\resizebox{\linewidth}{!}{%
\begin{tabular}{l c c c c c|c c}
\toprule
 & Exp. 1 & Exp. 2 & Exp. 3 & Exp. 4 & Exp. 5 & \makecell{Best Exp.\\Per LLM} & \makecell{Best LL\\Per LLM} \\
\midrule
GPT 4o & -- & -- & -- & $-979.47$ & $-969.73$ & Exp. 5 & $-969.73$ \\
GPT o4minihigh & -- & -- & $-999.72$ & $-987.79$ & $-981.47$ & Exp. 5 & $-981.47$ \\
GPT o3 & $-981.80$ & $-981.80$ & $-977.37$ & $-980.62$ & $-982.60$ & Exp. 3 & $-977.37$ \\
Claude 4 Opus & -- & -- & $-982.15$ & $-973.34$ & $-972.40$ & Exp. 5 & $-972.40$ \\
Claude 4 Sonnet & -- & -- & $-978.37$ & $-967.53$ & $-971.33$ & Exp. 4 & $-967.53$ \\
Mistral le Chat & -- & -- & $-982.69$ & $-1{,}030.97$ & $-974.06$ & Exp. 5 & $-974.06$ \\
DeepSeek V3.2 & -- & -- & $-970.47$ & $-980.95$ & $-956.68$ & Exp. 5 & $-956.68$ \\
Gemini 2.5 Flash & -- & -- & $-983.45$ & $-968.29$ & $-981.57$ & Exp. 4 & $-968.29$ \\
Gemma 3 & -- & -- & $-998.60$ & -- & -- & Exp. 3 & $-998.60$ \\
Llama 3 & -- & -- & -- & -- & $-991.85$ & Exp. 5 & $-991.85$ \\
Llama 4 Maverick & -- & -- & -- & -- & $-1{,}021.96$ & Exp. 5 & $-1{,}021.96$ \\
Llama 4 Scout & -- & -- & -- & -- & $-981.98$ & Exp. 5 & $-981.98$ \\
\midrule
Best LLM Per Exp. & GPT o3 & GPT o3 & DeepSeek V3.2 & Claude 4 S. & DeepSeek V3.2 & -- & -- \\
Best LL Per Exp. & $-981.80$ & $-981.80$ & $-970.47$ & $-967.53$ & $-956.68$ & -- & -- \\
\bottomrule
\end{tabular}%
}
\vspace{-2ex}
  {\scriptsize
    \begin{flushleft}
      Exp. 1: Full/ZS/Estimate, Exp. 2: Full/CoT/Estimate, Exp. 3: Full/ZS/Suggest, Exp. 4: Full/CoT/Suggest, Exp. 5: Limited/ZS/Suggest.
    \end{flushleft}
  }
\end{table}

\begin{table}[H]
\centering
\caption{Experiments Summary - AIC}
\label{tab:best-aic-exp}
\resizebox{\linewidth}{!}{%
\begin{tabular}{l c c c c c|c c}
\toprule
 & Exp. 1 & Exp. 2 & Exp. 3 & Exp. 4 & Exp. 5 & \makecell{Best Exp.\\Per LLM} & \makecell{Best AIC\\Per LLM} \\
\midrule
GPT 4o & -- & -- & -- & $1{,}992.94$ & $1{,}973.46$ & Exp. 5 & $1{,}973.46$ \\
GPT o4minihigh & -- & -- & $2{,}029.44$ & $1{,}993.59$ & $1{,}980.95$ & Exp. 5 & $1{,}980.95$ \\
GPT o3 & $1{,}977.61$ & $1{,}977.61$ & $1{,}984.74$ & $1{,}979.23$ & $1{,}991.20$ & Exp. 1 & $1{,}977.61$ \\
Claude 4 Opus & -- & -- & $1{,}980.31$ & $1{,}970.69$ & $1{,}960.81$ & Exp. 5 & $1{,}960.81$ \\
Claude 4 Sonnet & -- & -- & $1{,}976.75$ & $1{,}963.05$ & $1{,}972.15$ & Exp. 4 & $1{,}963.05$ \\
Mistral le chat & -- & -- & $1{,}977.39$ & $2{,}073.93$ & $1{,}970.13$ & Exp. 5 & $1{,}970.13$ \\
DeepSeek V3.2 & -- & -- & $1{,}960.93$ & $1{,}977.90$ & $1{,}945.36$ & Exp. 5 & $1{,}945.36$ \\
Gemini 2.5 Flash & -- & -- & $1{,}986.90$ & $1{,}966.59$ & $1{,}983.14$ & Exp. 4 & $1{,}966.59$ \\
Gemma 3 & -- & -- & $2{,}033.39$ & -- & -- & Exp. 3 & $2{,}033.39$ \\
Llama 3 & -- & -- & -- & -- & $2{,}021.70$ & Exp. 5 & $2{,}021.70$ \\
Llama 4 Maverick & -- & -- & -- & -- & $2{,}057.92$ & Exp. 5 & $2{,}057.92$ \\
Llama 4 Scout & -- & -- & -- & -- & $1{,}997.96$ & Exp. 5 & $1{,}997.96$ \\
\midrule
Best LLM Per Exp. & GPT o3 & GPT o3 & DeepSeek V3.2 & Claude 4 S. & DeepSeek V3.2 & -- & -- \\
Best AIC Per Exp. & $1{,}977.61$ & $1{,}977.61$ & $1{,}960.93$ & $1{,}963.05$ & $1{,}945.36$ & -- & -- \\
\bottomrule
\end{tabular}%
}
\vspace{-2ex}
  {\scriptsize
    \begin{flushleft}
      Exp. 1: Full/ZS/Estimate, Exp. 2: Full/CoT/Estimate, Exp. 3: Full/ZS/Suggest, Exp. 4: Full/CoT/Suggest, Exp. 5: Limited/ZS/Suggest.
    \end{flushleft}
  }
\end{table}

\begin{table}[H]
\centering
\caption{Experiments Summary - BIC}
\label{tab:best-bic-exp}
\resizebox{\linewidth}{!}{%
\begin{tabular}{l c c c c c|c c}
\toprule
 & Exp. 1 & Exp. 2 & Exp. 3 & Exp. 4 & Exp. 5 & \makecell{Best Exp.\\Per LLM} & \makecell{Best BIC\\Per LLM} \\
\midrule
GPT 4o & -- & -- & $2{,}099.91$ & $2{,}076.37$ & $2{,}056.89$ & Exp. 5 & $2{,}056.89$ \\
GPT o4minihigh & -- & -- & $2{,}098.17$ & $2{,}037.76$ & $2{,}025.12$ & Exp. 5 & $2{,}025.12$ \\
GPT o3 & $2{,}011.96$ & $2{,}011.96$ & $2{,}058.36$ & $2{,}023.40$ & $2{,}055.00$ & Exp. 1 & $2{,}011.96$ \\
Claude 4 Opus & -- & -- & $2{,}019.57$ & $2{,}016.90$ & $2{,}000.07$ & Exp. 5 & $2{,}000.07$ \\
Claude 4 Sonnet & -- & -- & $2{,}025.82$ & $2{,}031.76$ & $2{,}040.85$ & Exp. 3 & $2{,}025.82$ \\
Mistral le Chat & -- & -- & $2{,}006.83$ & $2{,}103.38$ & $2{,}024.11$ & Exp. 3 & $2{,}006.83$ \\
DeepSeek V3.2 & -- & -- & $2{,}010.01$ & $2{,}017.16$ & $2{,}020.56$ & Exp. 3 & $2{,}010.01$ \\
Gemini 2.5 Flash & -- & -- & $2{,}035.97$ & $2{,}040.21$ & $2{,}032.22$ & Exp. 5 & $2{,}032.22$ \\
Gemma 3 & -- & -- & $2{,}103.38$ & -- & -- & Exp. 3 & $2{,}103.38$ \\
Llama 3 & -- & -- & -- & -- & $2{,}114.94$ & Exp. 5 & $2{,}114.94$ \\
Llama 4 Maverick & -- & -- & -- & -- & $2{,}092.28$ & Exp. 5 & $2{,}092.28$ \\
Llama 4 Scout & -- & -- & -- & -- & $2{,}081.39$ & Exp. 5 & $2{,}081.39$ \\
\midrule
Best LLM Per Exp. & GPT o3 & GPT o3 & le Chat & Claude 4 O. & Claude 4 O. & -- & -- \\
Best BIC Per Exp. & $2{,}011.96$ & $2{,}011.96$ & $2{,}006.83$ & $2{,}016.90$ & $2{,}000.07$ & -- & -- \\
\bottomrule
\end{tabular}%
}
\vspace{-2ex}
  {\scriptsize
    \begin{flushleft}
      Exp. 1: Full/ZS/Estimate, Exp. 2: Full/CoT/Estimate, Exp. 3: Full/ZS/Suggest, Exp. 4: Full/CoT/Suggest, Exp. 5: Limited/ZS/Suggest.
    \end{flushleft}
  }
\end{table}

\noindent Finally, analysing the best-performing specifications in terms of BIC, which imposes a stronger penalty on model complexity than AIC, we observe some notable shifts (Table ~\ref{tab:best-bic-exp}). Claude 4 Opus generated the specification with the lowest BIC overall ($2{,}000.07$) in Experiment 5 (limited information), as well as across all experiments. This result reinforces earlier findings and underscores Claude’s ability in generating high-quality specifications under under limited information conditions. Claude 4 Opus also achieved the lowest BIC in Experiment 4 ($2{,}016.90$), while Mistral le Chat performed best in Experiment 3 ($2{,}006.83$). Four LLMs (GPT 4o, GPT o4-mini-high, Claude 4 Opus, and Gemeni 2.5 Flash) achieved their best BIC outcomes in Experiment 5. These results further support the emerging pattern that simplified/limited inputs may encourage LLMs to focus more on utility specification and behavioural reasoning, rather than being distracted by more complex data analysis.

\subsection{Overall Evaluation of LLM Performance}
\label{results_2}
To complement the experiment-level analysis, this subsection provides a broader evaluation of each LLM’s overall performance across all configurations. Table \ref{tab:llm-metrics} summarizes the performance and characteristics of the utility specifications generated by each LLM across all experiments, offering a broader view of their modelling capabilities. Note that open-weight models (Gemma 3  and Llama variants) are excluded from this overall evaluation as they produced few valid specifications across expriments (Gemma 3 only in Exp. 3 and Llama variants only in Exp. 5), which renders average-based comparisons unreliable.

\begin{table}[H]
\centering
\caption{Overall Evaluation of Generated Specifications by LLM}
\label{tab:llm-metrics}
\resizebox{\linewidth}{!}{%
\begin{tabular}{l c c c c c c c c c c}
\toprule
LLM & \makecell{Av. Nb\\of Spec.} & \makecell{Models\\Converged} & \makecell{Av. Nb\\of Vars} & \makecell{Av. Nb\\of Params} & \makecell{Generic\\Params} & \makecell{Alt-Spec.\\Params} & \makecell{ASC\\Included} & \makecell{Av. Nb of\\Socioeconomics} & \makecell{Av. Nb of\\Transformations} & \makecell{Av. Nb of\\Interactions} \\
\midrule
ChatGPT 4o & $2.58$ & $100\%$ & $4.00$ & $10.75$ & $23\%$ & $77\%$ & $58\%$  & $0.50$ & $0.50$ & $0.42$ \\ 
ChatGPT o4-mini-high & $3.00$ & $100\%$ & $3.53$ & $7.80$  & $47\%$ & $53\%$ & $93\%$ & $0.33$ & $0.73$ & $0.40$ \\
ChatGPT o3 & $2.77$ & $100\%$ & $3.68$ & $7.18$  & $47\%$ & $53\%$ & $100\%$ & $0.27$ & $0.59$ & $0.59$ \\
Claude 4 Opus & $3.68$ & $89\%$  & $3.68$ & $7.74$  & $43\%$ & $57\%$ & $100\%$ & $0.37$ & $0.26$ & $0.53$ \\
Claude 4 Sonnet & $3.50$ & $100\%$ & $4.67$ & $9.33$  & $36\%$ & $64\%$ & $100\%$ & $1.22$ & $0.33$ & $0.72$ \\
Mistral le Chat & $3.47$ & $88\%$  & $3.06$ & $6.71$  & $45\%$ & $55\%$ & $100\%$ & $0.29$ & $0.24$ & $0.41$ \\
DeepSeek V3.2 & $3.19$ & $94\%$  & $3.56$ & $9.50$  & $34\%$ & $66\%$ & $100\%$  & $0.12$ & $0.12$ & $0.88$ \\
Gemini 2.5 Flash & $2.93$ & $100\%$ & $3.50$ & $10.71$ & $32\%$ & $68\%$ & $100\%$ & $0.36$ & $0.43$ & $0.43$ \\
\bottomrule
\end{tabular}%
}
\end{table}

\noindent Across all experiments, Claude variants generated on average more specifications compared to GPT, DeepSeek, and Gemini variants, as discussed previously. Specifically, Claude 4 Opus generated on average the highest number of specifications (3.68), though only 86\% of these successfully converged during estimation. Claude 4 Sonnet ranked second, generating an average of 3.50 specifications, closely followed by Mistral le Chat with 3.47. In addition to Claude 4 Opus, both Mistral le Chat and DeepSeek V3.2 had some convergence challenges, with convergence rates of 88\% and 94\%, respectively. In contrast, all GPT variants, Gemini 2.5 Flash, and Claude 4 Sonnet achieved a perfect 100\% convergence rate, demonstrating greater robustness in suggesting estimable specifications.\\

\noindent With respect to model complexity, Claude 4 Sonnet stands out as the LLM model that generated, on average, the most complex specifications. This included the highest numbers of variables (4.67), a relatively high number of parameters (9.33), the highest number of socioeconomic variables (1.22), and the second highest number of interactions between socioeconomic variables and attributes (0.72). Conversely, GPT-o3, despite being OpenAI's most advanced reasoning model at the time of writing this paper, proposed relatively simpler specifications, with the second lowest average number of parameters (7.18) and few socioeconomic variables (0.27). This outcomes is likely influenced by the fact that GPT-o3 had to allocate its reasoning capabilities both to suggesting specifications and estimating them in Experiments 1 and 2, rather than allocating all its resources to generating more complex specifications. \\

\noindent While GPT 4o and Gemini 2.5 suggest specifications with the highest number of parameters (10.75 and 10.71, respectively), GPT 4o struggled with including alternative-specific constants, with only 58\% of its specifications incorporating them. GPT o4-minihigh encountered a similar issue, with ASCs included in 93\% of its specifications. However, GPT 4o had a strong inclination towards modelling behavioural heterogeneity across alternatives, with 77\% of its parameters specified in alternative-specific form.\\ 

\noindent In terms of representing systematic observed heterogeneity, Claude 4 Sonnet again led with the highest average use of socioeconomic variables (1.22 per specification), whereas Deepseek incorporated these variables far less frequently (0.12 on average). This shows differences across LLM families in their capability for capturing behavioural heterogeneity across different population segments. The use of functional transformations, such as logarithmic, Box-Cox, and piece-wise forms, was generally limited across all LLMs. However, GPT-o4-mini-high (0.73) and GPT-o3 (0.59) applied them slightly more often, suggesting a better tendency toward including non-linear relationships in their utility functions. \\

\begin{figure}[H]
    \centering
    \includegraphics[origin=c, width = 1.0 \textwidth
   ]{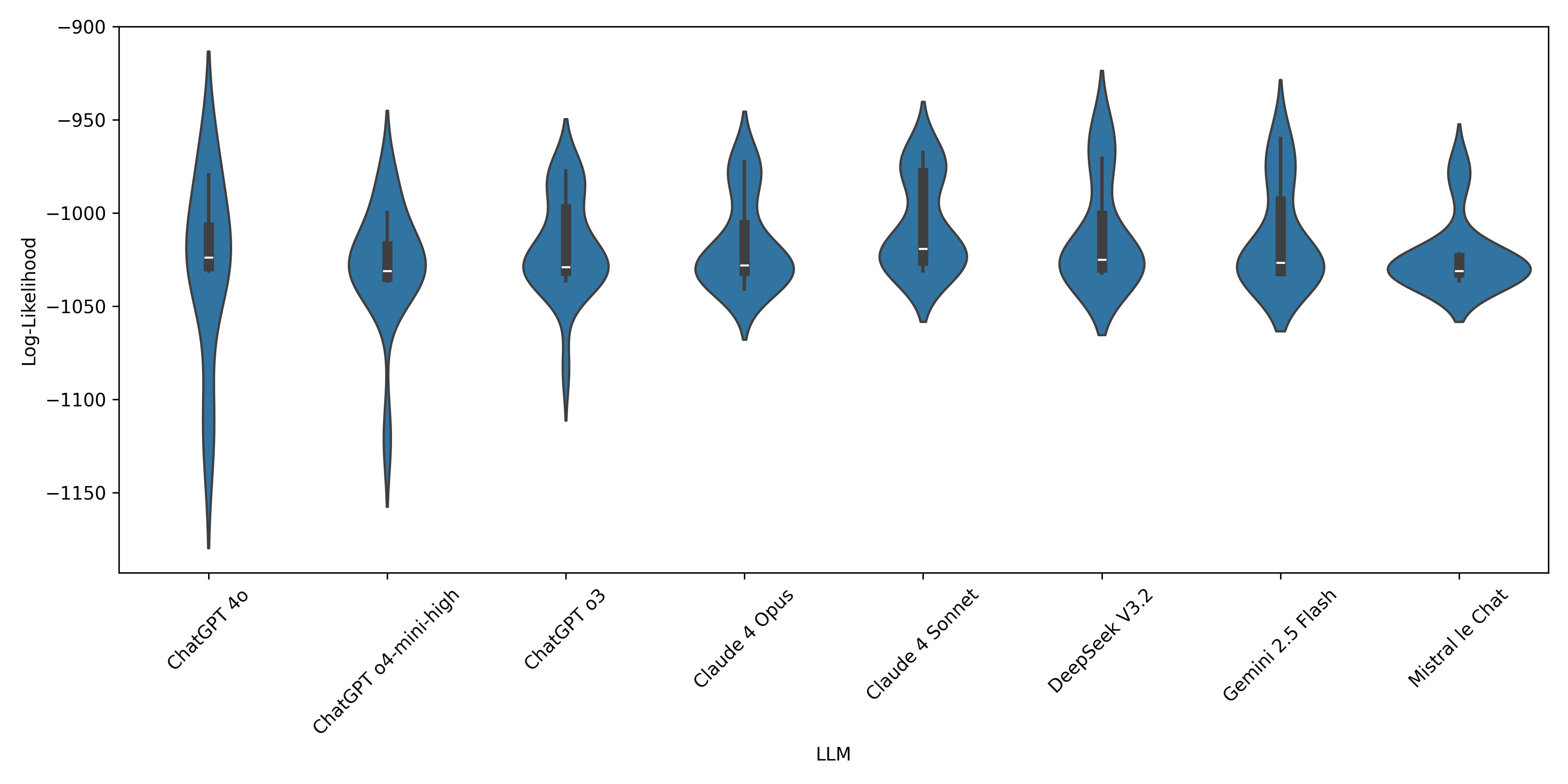}
    \caption{Log-Likelihood Distributions of Generated Specifications across LLMs}
    \label{ll-llms}
\end{figure}

\begin{figure}[H]
    \centering
    \includegraphics[origin=c, width = 1.0 \textwidth
   ]{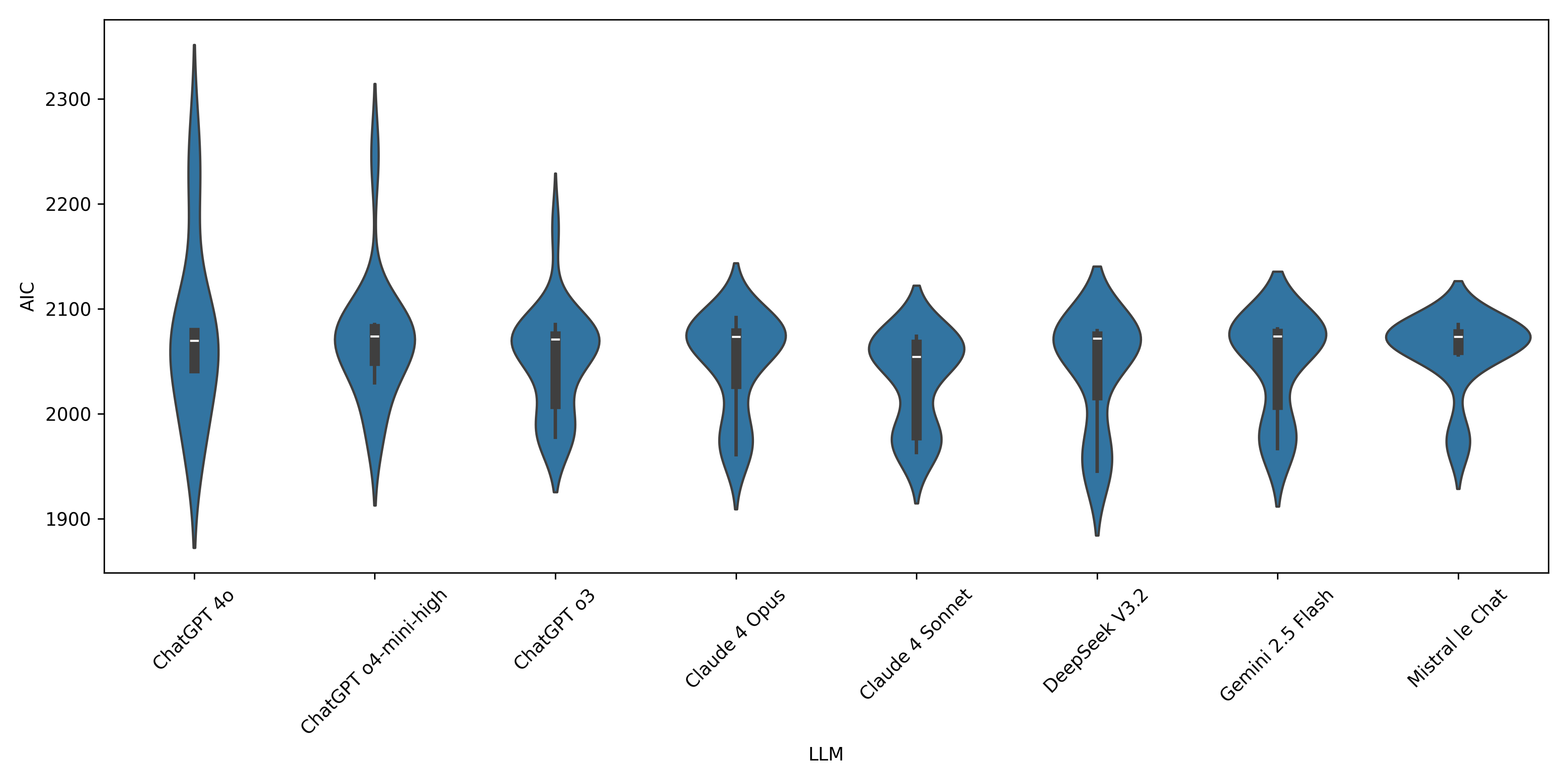}
    \caption{AIC Distributions of Generated Specifications across LLMs}
    \label{aic-llms}
\end{figure}

\begin{figure}[H]
    \centering
    \includegraphics[origin=c, width = 1.0 \textwidth
   ]{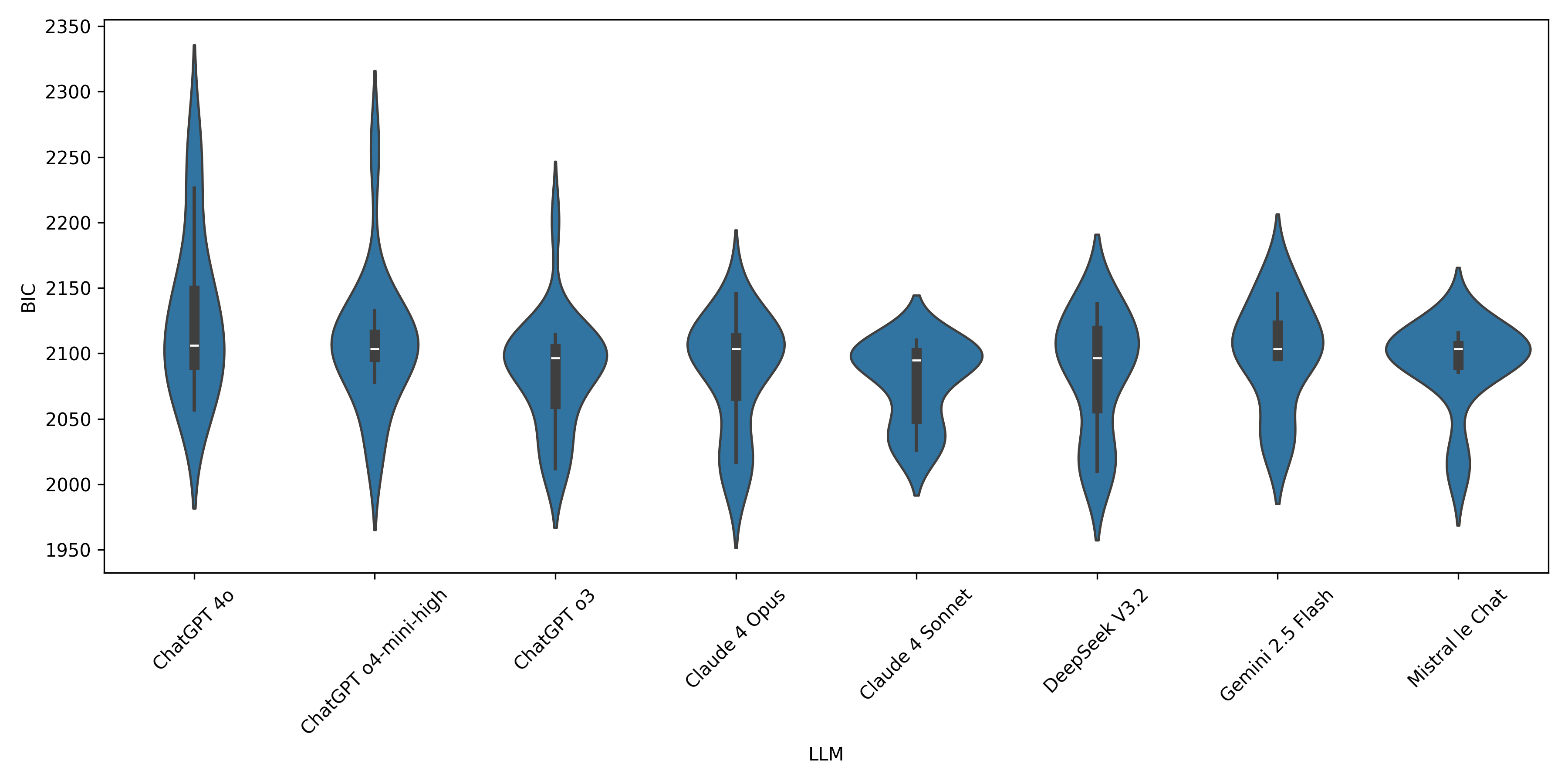}
    \caption{BIC Distributions of Generated Specifications across LLMs}
    \label{bic-llms}
\end{figure}

\noindent While the summary of utility specifications for each LLM provides a static overview of model complexity and specification trends, the distribution of modelling outcomes also provides a better understanding of the LLMs' overall performance. Figures \ref{ll-llms}, \ref{aic-llms}, and \ref{bic-llms} display the full distributions of LL, AIC, and BIC values of all specifications that did converge by each LLM.  These plots enable comparison not only in terms of central tendency, but also in the consistency and reliability of each LLM’s specification quality. Overall, the medians of the LL, AIC, and BIC distributions lie in a relatively narrow range across all LLMs, indicating that, on average, most models are capable of producing specifications with broadly comparable empirical performance. However, Claude 4 Sonnet has, by a slight margin, the uppermost position in the LL plot and the lowest position in the AIC and BIC plots, indicating that it generates, on average, the best-fitting and most efficient specifications. Differences emerge more clearly in terms of dispersion and tail behaviour. Claude variants, DeepSeek, Gemini, and Mistral display relatively compact distributions, implying stable performance and a more consistent balance between goodness-of-fit and model complexity. By contrast, ChatGPT 4o and ChatGPT o4-mini-high exhibit narrower widths around their medians and longer tails, implying a greater proportion of weaker specifications in terms of LL, AIC, and BIC.\\

\begin{figure}[H]
    \centering
    \includegraphics[origin=c, width = 1.0 \textwidth
   ]{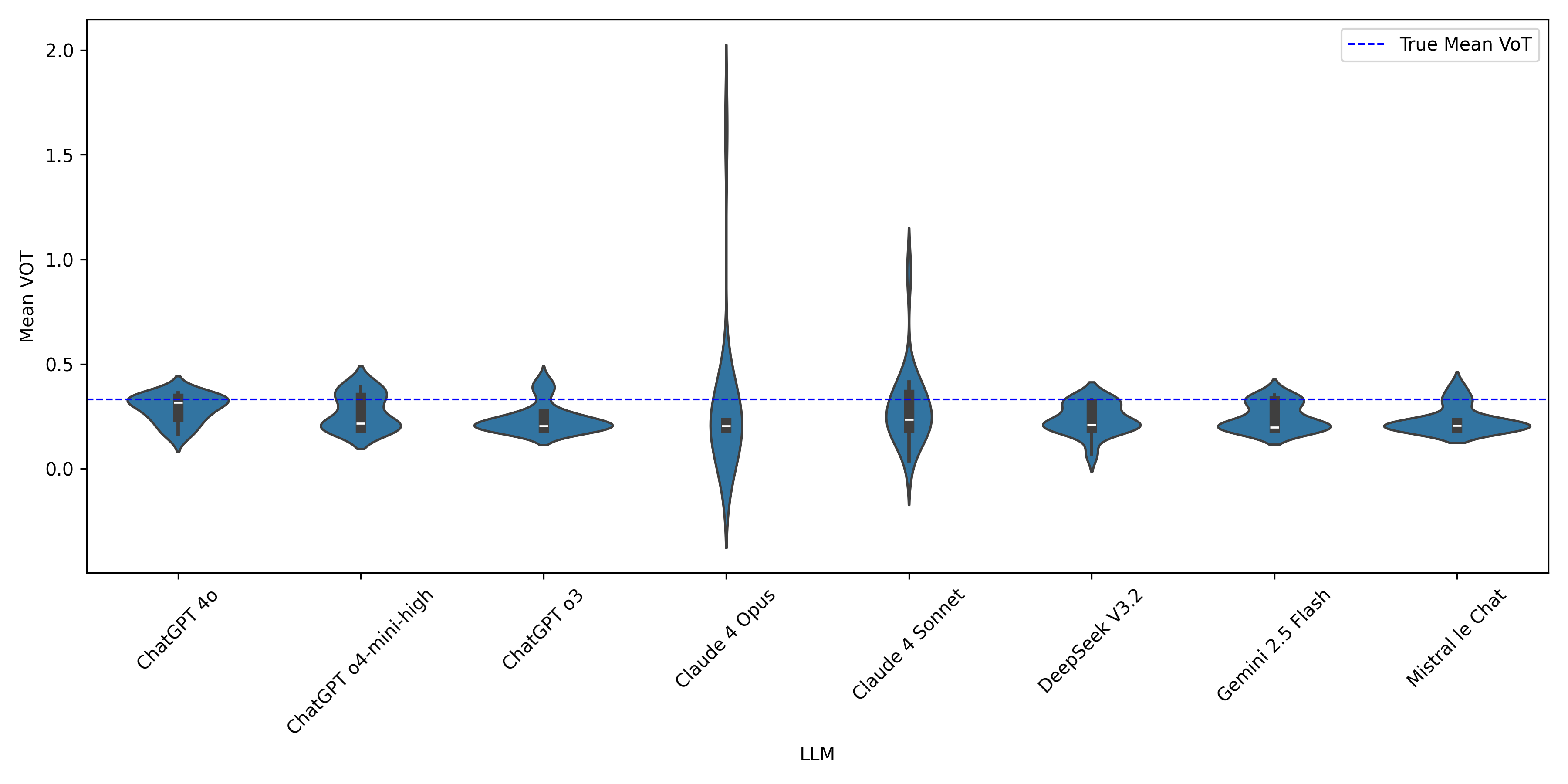}
    \caption{VOT Distributions of Generated Specifications across LLMs}
    \label{vot-llms}
\end{figure}

\noindent Finally, Figure \ref{vot-llms} presents the distribution of the mean Value of Time (VoT) estimates derived from all valid specifications generated by each LLM. For each specification, the VoT is calculated as the ratio of the travel-time coefficient to the cost coefficient, averaged across all alternatives. The blue dashed line indicates the true mean VoT from the synthetic dataset. Most LLMs yielded VoT estimates closely centred around the true value with narrow distributions, indicating that the generated specifications were effective at accurately capturing the trade-off between time and cost in individuals' decision-making. However, Claude 4 Opus and Claude 4 Sonnet show greater variability, with some specifications yielding high values. However, in some of these specifications, cost coefficients were statistically insignificant, resulting in unreliable VoT estimates.

\subsection{Variability of LLM Outputs Across Repeated Prompting}
\label{repeat_promp}

Large language models generate outputs probabilistically, and repeated executions of the same prompt may therefore yield different results. To assess the variability and reliability of LLM-generated specifications under stochasticity, we conducted multiple replications for Experiment 5. This experiment was selected for replication as it yielded several of the best results (Tables \ref{tab:best-ll-exp}, \ref{tab:best-aic-exp}, \ref{tab:best-bic-exp}), but also because it provides the most comparable setting across model architectures.\\

\begin{figure}[H]
    \centering
    \includegraphics[origin=c, width = 1.0 \textwidth
   ]{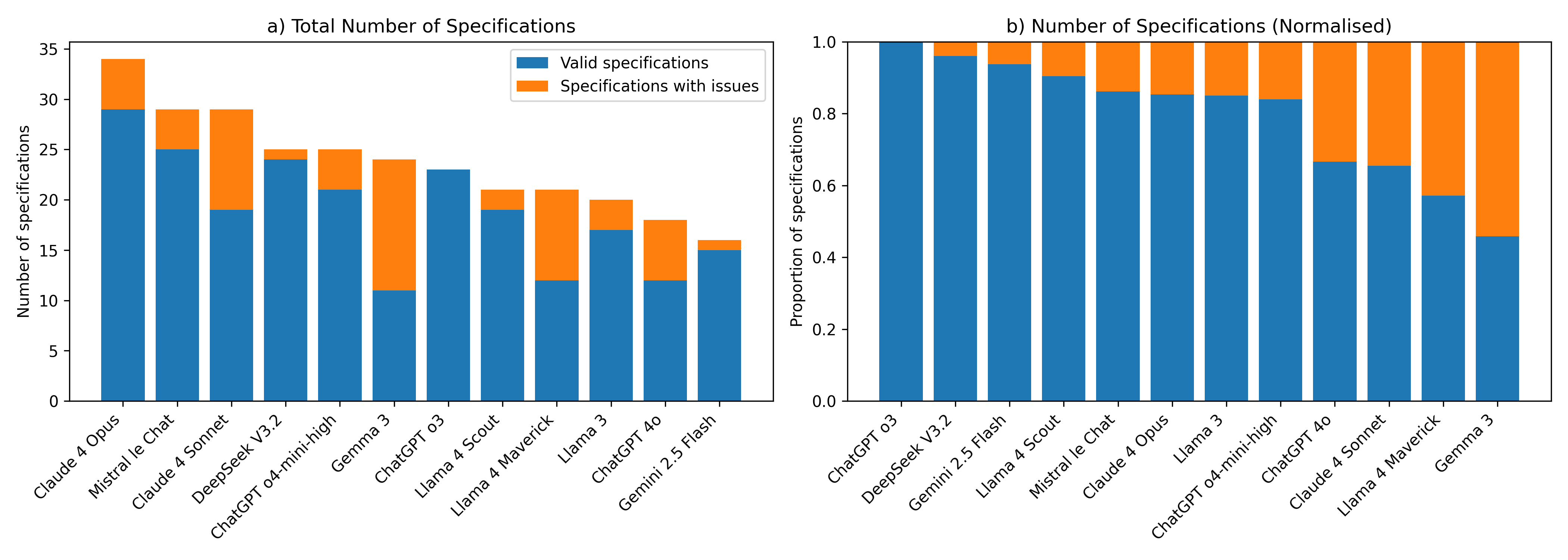}
    \caption{Number and validity of specifications across five replications (Experiment 5). a) Total number of model specifications generated by each LLM across five independent runs of Experiment 5, distinguishing between valid specifications and specifications with issues (e.g., non-convergence or behavioural implausibility). b) Normalised proportions of valid and  specifications with issues for each LLM, highlighting relative reliability independently of the total number of generated specifications.}
    \label{fig:exp5_5runs}
\end{figure}

\noindent In particular, Experiment 5 is run in a non-agentic configuration with limited information and ``Suggest" goal. This setting restricts all models to the same capabilities (text and code generation and reasoning), without relying on external tools or environments. For each LLM, the prompt was executed five times under fixed generation settings. All resulting specifications were independently implemented and estimated following the same procedure as in the single-run experiments. Figure \ref{fig:exp5_5runs} summarises the total number of specifications generated by each LLM across the five runs, distinguishing between valid specifications and specifications with issues (e.g. non-convergence or behavioural implausibility). It highlights differences between LLMs in terms of reliability that are consistent with patterns previously observed in the single-run analysis (Table \ref{tab:llm-metrics}). Specifically, Claude variants and Mistral le Chat generated the highest number of specifications with low shares of problematic specifications. GPT-o3, DeepSeek V3.2, and Gemini 2.5 Flash exhibited high proportions of valid specifications. Open-weight models (Llama and Gemma), in contrast, showed both lower generation capabilities and higher proportions of specifications with issues, even under repeated execution.\\

\begin{figure}[H]
    \centering
    \includegraphics[origin=c, width = 1.0 \textwidth
   ]{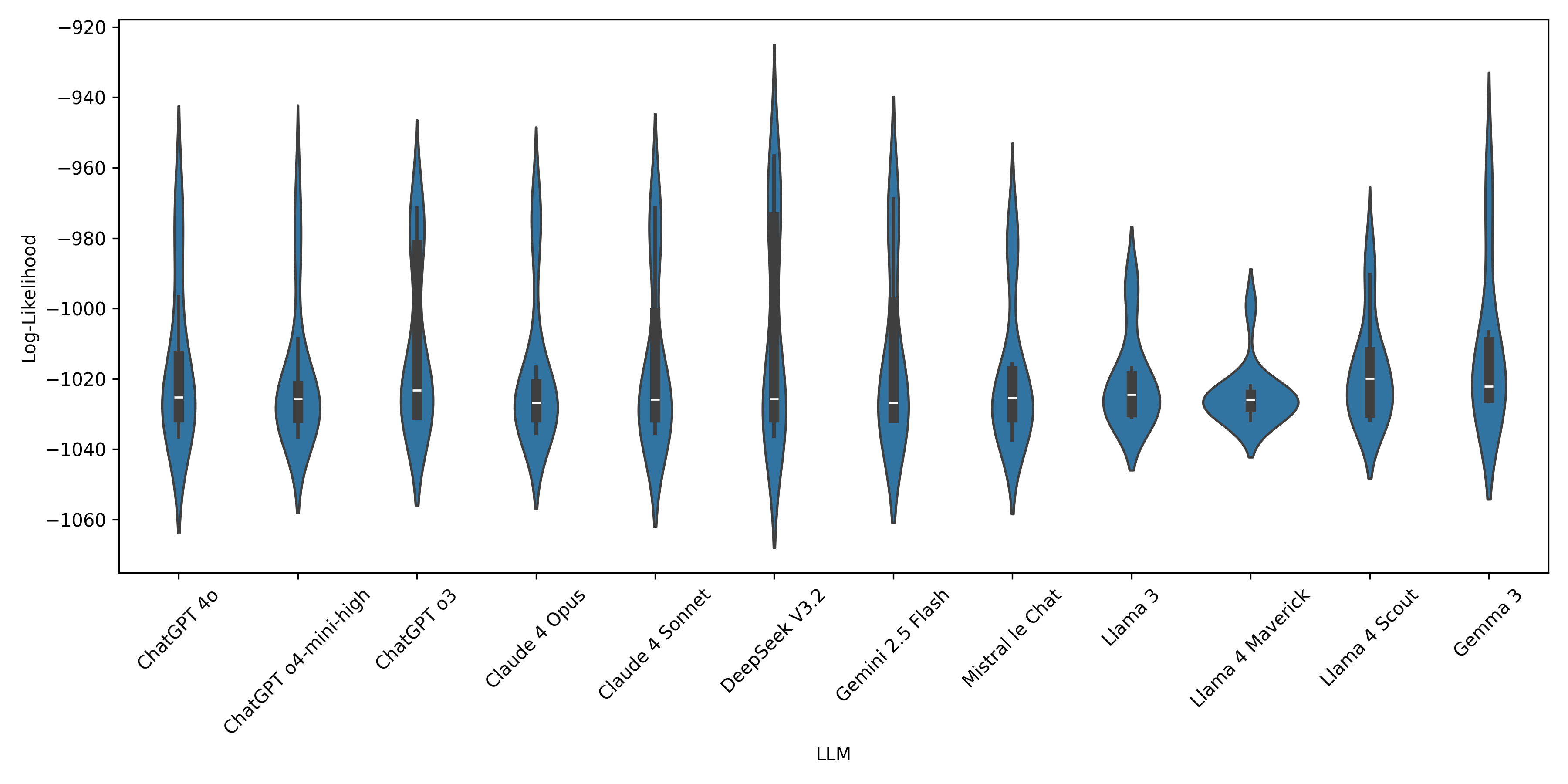}
    \caption{Distribution of log-likelihood values across five replications (Experiment 5)}
    \label{fig:exp5_5runs_LL}
\end{figure}

\noindent Figure \ref{fig:exp5_5runs_LL} presents the log-likelihood distributions of all valid specifications generated across the five runs for each LLM. While within-model variability in model fit is observed, as expected given the probabilistic nature of LLM outputs, the relative ordering/ranking of LLMs in terms of model performance remains broadly stable across replications. Closed-weight models that performed well in the single-run analysis continued to do so on average, while open-weight models that generated weaker specifications remained consistently less competitive.

\section{Conclusion}
\label{section:con}
This study provides a systematic evaluation of how current LLMs can assist in the specification and, where technically feasible, the estimation of MNL models. Our aim was twofold: to understand \textbf{how modelling goals, prompting strategies, and information conditions affect model quality}, and to assess \textbf{which LLMs currently perform best in discrete choice modelling tasks}. To do so, we designed five experimental configurations that vary along three key dimensions — \textbf{prompting strategy} (zero-shot vs. chain-of-thought), \textbf{information availability} (full data vs. data dictionary), and \textbf{modelling task} (suggestion vs. estimation). Moreover, we benchmarked twelve versions of six leading LLM families: ChatGPT, Claude, DeepSeek, Gemini, Gemma, Llama, and Mistral. Across all configurations, we assessed the empirical performance, behavioural plausibility, and complexity of each utility specification generated. \\

\noindent Our findings offer three main insights. First, prompt structure plays a critical role in shaping specification quality. Structured chain-of-thought prompts led, on average, to higher-quality and better-performing specifications than unstructured zero-shot prompts. Second, limited access to raw data did not reduce—and in some cases improved—LLM performance. This suggests that constraining input complexity may allow LLMs to allocate more of their reasoning and computational resources toward behavioural and theoretical reasoning rather than parsing data structure. Third, model performance vary significantly across LLM families as a function of their architectural configurations and access to external computational tools. End-to-end estimation was only feasible for agentic tool-augmented LLMs. Specifically, GPT-o3, operating in an agentic setting, was uniquely capable of estimating its own models through executable, verifiable code, demonstrating the potential for partial automation of the entire modelling pipeline. In contrast, non-agentic LLMs lacked the necessary execution capabilities, resulting in hallucinated or non-reproducible outputs. These findings demonstrate the potential of incorporating LLMs into different parts of the choice modelling workflow, with non-agentic LLMs supporting specification generation and agentic LLMs supporting estimation).\\

\noindent However, several limitations remain. Some LLM-generated specifications lacked alternative-specific constants, failed to converge when independently estimated, or produced counter-intuitive parameter signs. These limitations underscore the continued necessity of expert oversight in model development, interpretation, and validation.\\

\noindent Moreover, open-weight models evaluated in this study (Llama and Gemma variants) exhibited significant limitations when used for end-to-end choice modelling workflows in out-of-the-box manner, without fine-tuning, and under practical deployment constraints. Unlike proprietary (closed-weight) models, which were accessed through managed interfaces, these models were executed through inference APIs, which introduces additional technical complexity and may impose computational and memory constraints. Empirically, both models failed frequently to generate behaviourally plausible utility specifications or produced specifications that did not converge when estimated. Gemma produced usable outputs only in one experimental configuration (Experiment 3), and only after invalid specifications were manually filtered out. Llama had interface limitations that prevented file uploads at the time the experiments were conducted. As a result, it succeeded in generating estimable specifications only in Experiment 5, where inputs were limited to a prompt and data dictionary. Across experiments, these models tended to return very few specifications due to limited output length and often lacked access to, or understanding of, the database structure. Importantly, these findings should not be interpreted as inherent limitations of open-weight LLM architectures in general. Rather, they reflect the performance of the particular model versions evaluated here, when used without task-specific fine-tuning and under realistic deployment conditions. Fine-tuning or tighter integration within hybrid systems may plausibly improve their performance. Accordingly, while the open-weight models considered here remain unsuitable for reliable end-to-end automation in discrete choice modelling, they may still play a valuable role within hybrid workflows that leverage their strengths in language understanding, reasoning, and prompt-based guidance, without relying on them for full automation (see \citep{cao2024survey} for a detailed taxonomy of LLM-enhanced  reinforcement learning framework, which could be extended to discrete choice modelling \citep{nova2025improving}).\\

\noindent In summary, our findings offer several practical insights for how to best use LLMs in specifying and estimating MNL models:

\begin{itemize}
    \item \textbf{Prompt structure matters:} Structured prompting (chain-of-thought) significantly enhances specification quality compared to unstructured(zero-shot) prompts.
    \item \textbf{Less can be more:} Limiting LLM access to raw detailed data access can sometimes lead to stronger theoretical reasoning by LLMs, enhancing their performance in generating better utility specifications.
    \item \textbf{End-to-end automation is configuration-dependent:} Only agentic LLMs are capable of both proposing theoretically sound MNL specifications and correctly estimating them, by generating valid Python code, executing it, and returning verifiable log-likelihood values and parameter estimates. In our evaluation, this was observed exclusively for GPT-o3.
\end{itemize}

\noindent Regarding model-specific performance, our findings suggest the following:
\begin{itemize}
    \item \textbf{Claude models} frequently explored the modelling space more extensively by generating a larger number of candidate specifications. Under structured prompting (CoT), Claude 4 Sonnet produced the best-performing specification in the full-information setting.
    \item \textbf{GPT variants} showed robustness in convergence and consistency in generating reliable specifications, although GPT 4o omitted ASCs from a larger share of its specifications.
    \item \textbf{Gemini 2.5 Flash} also showed robustness in convergence and consistency in generating reliable specifications, albeit generally less complex than Claude's.
    \item \textbf{DeepSeek V3.2} generated some of the best specifications overall. It achieved the best fit in multiple experiments and produced the best LL across all configurations in the limited information setting.
    \item \textbf{Open-weight models} (Llama and Gemma) underperformed overall. They produced very few valid specifications, and even those were rarely competitive with the outputs of closed-weight models in terms of LL, AIC, and BIC.
\end{itemize}

\noindent Finally, this work illustrates both the potential and the current boundaries of LLMs in discrete choice modelling. While these models can assist in utility specification, they should be viewed as support tools rather than autonomous agents. Future research should explore how to refine prompt engineering to improve specifications quality and build hybrid workflows that effectively combine human expertise with LLM-driven support.

\clearpage
\section*{Acknowledgement}
The authors acknowledge the financial support by the European Research Council through the advanced grant 101020940-SYNERGY.

\clearpage
\section*{Appendix}
\begin{algorithm}\label{API}
\caption{GROQ API Completion Flow}
\begin{lstlisting}[style=mystyle, language=Python]
# Initialize Groq client with API key
from groq import Groq
import os 

# Get a key at www.groq.com
key = user_key # copy from groq

# Select prompt 
prompt = prompt_zero_shot_suggest # or prompt_CoT_suggest 

client = Groq(api_key=os.environ.get("GROQ_API_KEY", key),)

# Configure output location
output_dir = "llama_4_scout_17b"  # Options:
# output_dir = "llama_4_maverick_17b"
# output_dir = "llama_3_70b_8192"

file_name = "Z_suggest.txt"  # or CoT_suggest.txt
full_path = os.path.join(output_dir, file_name)

# Generate completion
llama_output = ""
completion = client.chat.completions.create(    
    model="meta-llama/llama-4-scout-17b-16e-instruct",
    
    # You may find other models such as:
    # model="meta-llama/llama-4-maverick-17b-128e-instruct",
    # model="llama3-70b-8192",
    
    messages=[{"content": prompt}],
    temperature=1.2,       # Control creativity
    top_p=0.95,            # Control diversity
    max_completion_tokens=8192,  # Max response length
    stop=None,             # No early stopping  )

# Process and save output
for chunk in completion:
    content = chunk.choices[0].delta.content or ""
    llama_output += content
with open(full_path, "w", encoding="utf-8") as f:
    f.write(llama_output)

\end{lstlisting}
\end{algorithm}

\begin{algorithm}\label{API2}
\caption{OpenRouter API Completion Flow}
\begin{lstlisting}[style=mystyle, language=Python]
# Initialize OpenRoute key
from openai import OpenAI
import os, requests, time, json

# Get a key at https://openrouter.ai
client = OpenAI(base_url="https://openrouter.ai/api/v1", api_key=api_key,timeout=180.0,)

# --- Config ---
prompt, num_runs = prompt_zero_shot_suggest, 5
output_dirs = ["llama4_scout"]
models = ["meta-llama/llama-4-scout"]
base_file_name = "zero_shot_suggest.txt"

for output_dir, model in zip(output_dirs, models):    
    os.makedirs(output_dir, exist_ok=True)
    EXTRA_HEADERS = {"HTTP-Referer": "http://localhost", "X-Title": "gemma3-experiment"}

    def chat_once():
        """Raw HTTP call (useful when SDK fails to decode JSON)."""
        url = "https://openrouter.ai/api/v1/chat/completions"  
        payload = {"model": model,
            "messages": [{"role": "system", "content": "You are a helpful assistant."}, {"role": "user", "content": prompt},],
            "temperature": 1.2,
            "top_p": 0.95,
            "max_tokens": 16384,}
        headers = {"Authorization": f"Bearer {api_key}",
            "Content-Type": "application/json",
            **EXTRA_HEADERS,}
        r = requests.post(url, headers=headers, json=payload, timeout=120)
        return r.json()["choices"][0]["message"]["content"] or ""

    for run_idx in range(1, num_runs + 1):
        run_name = f"run_{run_idx}"
        out_path = os.path.join(output_dir, f"{run_name}_{base_file_name}")
        text = chat_once

        with open(out_path, "w", encoding="utf-8") as f:
            f.write(text)
        time.sleep(1.0)  

\end{lstlisting}
\end{algorithm}

\clearpage
\bibliography{sources/references}
\end{document}